\documentclass[conference]{IEEEtran}

\usepackage{url}
\usepackage{soul}
\usepackage{pifont}
\usepackage{xspace}
\usepackage{amsthm}
\usepackage{diagbox}
\usepackage{booktabs}
\usepackage{hyperref}
\usepackage{makecell}
\usepackage{multirow}
\usepackage{graphicx}
\usepackage{ragged2e}
\usepackage{algorithm}
\usepackage{enumerate}
\usepackage{todonotes}
\usepackage{tablefootnote}
\usepackage{algpseudocode}
\usepackage{color, xcolor}
\usepackage{amsmath, amsfonts}

\newcommand{\ie}{\textit{i.e.,}\xspace}
\newcommand{\eg}{\textit{e.g.,}\xspace}
\newcommand{\etc}{\textit{etc.}\xspace}
\newcommand{\etal}{\textit{et al.}\xspace}

\newcommand{\figref}[1]{Fig.~\ref{#1}\xspace}
\newcommand{\tabref}[1]{TABLE~\ref{#1}\xspace}
\newcommand{\secref}[1]{Section~\ref{#1}\xspace}
\newcommand{\equref}[1]{Equation~\ref{#1}\xspace}

\newcommand{\toolname}{{\sc \textbf{ReCoDe}}\xspace}
\newcommand{\classifier}{{\sc \textbf{Classifier}}\xspace}
\newcommand{\augmentor}{{\sc \textbf{Augmentor}}\xspace}
\newcommand{\decomposer}{{\sc \textbf{Decomposer}}\xspace}
\newcommand{\detector}{{\sc \textbf{Detector}}\xspace}

\newcommand{\cs}{{\sc \textit{Classifying Stage}}\xspace}
\newcommand{\ds}{{\sc \textit{Detecting Stage}}\xspace}

\newcommand{\bert}{{\sc \textbf{ReCoDe-BERT}}\xspace}

\newcommand{\func}{\textit{Functional Defect}\xspace}
\newcommand{\crash}{\textit{Crash}\xspace}
\newcommand{\laypro}{\textit{Layout Problem}\xspace}
\newcommand{\dispro}{\textit{Display Problem}\xspace}
\newcommand{\neterr}{\textit{Network Error}\xspace}
\newcommand{\nullsr}{\textit{Null Screen}\xspace}
\newcommand{\perfp}{\textit{Performance Problem}\xspace}
\newcommand{\errorp}{\textit{Error Prompt}\xspace}
\newcommand{\garerr}{\textit{Garbled Error}\xspace}
\newcommand{\tranpr}{\textit{Transition Problem}\xspace}

\newcommand{\typeNum}{{10}\xspace}
\newcommand{\dataNum}{{22,720}\xspace}
\newcommand{\conNum}{{4,105}\xspace}
\newcommand{\conRate}{{18.07\%}\xspace}

\newcommand{\tabincell}[2]{\begin{tabular}{@{}#1@{}}#2\end{tabular}}

\definecolor{bgcolor}{RGB}{235, 235, 235}
\newcommand{\bgg}[1]{\begingroup\sethlcolor{bgcolor}\textcolor{black}{\hl
{\textbf{#1}}}\endgroup}

\newcommand{\vv}[1]{#1}
\newcommand{\vvvvv}[1]{#1}

\begin{document}

\title{Mobile App Crowdsourced Test Report Consistency Detection via Deep Image-and-Text Fusion Understanding}

\author{
\IEEEauthorblockN{Shengcheng Yu, Chunrong Fang*, Quanjun Zhang, Zhihao Cao, Yexiao Yun, Zhenfei Cao, Kai Mei, Zhenyu Chen}
\IEEEauthorblockA{State Key Laboratory for Novel Software Technology, Nanjing University, Nanjing, China \\ Corresponding Author: fangchunrong@nju.edu.cn}
}

\maketitle

\begin{abstract}

Crowdsourced testing, as a distinct testing paradigm, has attracted much attention in software testing, especially in mobile application (app) testing field. Compared with in-house testing, crowdsourced testing shows superiority with the diverse testing environments when faced with the mobile testing fragmentation problem. However, crowdsourced testing also encounters the low-quality test report problem caused by unprofessional crowdworkers involved with different expertise. In order to handle the submitted reports of uneven quality, app developers have to distinguish high-quality reports from \vv{low-quality} ones to help the bug inspection. One kind of typical \vv{low-quality} test report is inconsistent test reports, which means the textual descriptions are not focusing on the attached \vv{bug-occurring} screenshots. According to our empirical survey, only \conRate crowdsourced test reports are consistent. Inconsistent reports cause waste on mobile app testing.

To solve the inconsistency problem, we propose \toolname to detect the consistency of crowdsourced test reports via deep image-and-text fusion understanding. \toolname is a two-stage approach that first classifies the reports based on textual descriptions into different categories according to the bug feature. In the second stage, \toolname has a deep understanding \vv{of} the GUI image features of the app screenshots and then applies different strategies to handle different types of bugs to detect the consistency of the crowdsourced test reports. We conduct an experiment on a dataset with over 22k test reports to evaluate \toolname, and the results show the effectiveness of \toolname \vv{ in detecting the consistency} of crowdsourced test reports. \vv{Besides, a user study is conducted to prove the practical value of \toolname in effectively helping app developers improve the efficiency of reviewing the crowdsourced test reports.}

\end{abstract}


\section{Introduction}

Crowdsourced testing has become one of the mainstream testing paradigms, and it is famous for its openness \cite{feng2016multi} \cite{yu2021prioritize}. The advantage of crowdsourced testing is obvious. Crowdsourced testing makes it possible for the applications (app) under test to run on different testing environments \cite{gao2019successes}, \ie device model, manufacturer, operating system, screen resolution, \etc Therefore, both functionality and compatibility can be fully tested \cite{feng2015test}. 

\vv{However, the openness of crowdsourced testing also brings some problems. The first one is that crowdsourced testing always involve a large number of participants (\ie crowdworkers), who will submit large number of crowdsourced test reports.} To submit a crowdsourced test report, crowdworkers are required to describe the bug occurring context and unexpected app behaviors, which are always mixed in a report \cite{yu2021prioritize}. App screenshots are necessary to assist in presenting the bugs \cite{feng2016multi}. Usually, crowdworkers are encouraged to attach one app screenshot to each test report, illustrating the bug occurrence. As a consequence, reviewing such reports brings extra workload and risks \cite{wang2019images} to app developers. \vv{To alleviate this problem, some existing studies have done a lot of work on the collected crowdsourced test reports to reduce the bug amount, like duplicate detection \cite{wang2019images}, report prioritization \cite{feng2016multi} \cite{yu2021prioritize}, report clustering, \etc} 

\vv{Another important problem brought by the openness of crowdsourced testing is the quality control of crowdsourced test reports.} Crowdsourced testing is different from in-house testing, the expertise of crowdworkers is widely ranged and thus the quality of the reports they submit cannot be guaranteed \cite{yu2019crowdsourced}. 
\vv{During the bug inspection and reproduction process, app developers need to refer to the app screenshots to identify which app activity has a problem, because app screenshots and textual descriptions can provide information from different perspectives to form a complete bug context. If app screenshots and textual descriptions are not consistent with each other, app developers have to distinguish which information is reliable, and thus confusion can be caused. In other words, consistent reports can help app developers more effectively inspect and reproduce bugs.

As a basic requirement to form a high-quality crowdsourced test report, app screenshots and textual descriptions are supposed to match each other to help depict the bugs \cite{gao2019successes}, while in inconsistent reports, the app screenshots and textual descriptions do not match each other, which is confusing for app developers to figure out which information can be referred to. The mismatch between the app screenshots and textual descriptions also weakens the effect of the mutual confirmation of app screenshots and textual descriptions, making app developers have to spend more time in such inconsistent reports. It is a great burden for app developers to review the mixed high-quality and low-quality reports. Consequently, in most cases, such reports can only be discarded.}

\vv{In short, consistent reports can effectively help app developers more efficiently review and reproduce bugs, while inconsistent bug reports will lead to confusion when app developers both refer to app screenshots and textual descriptions when reviewing and reproducing bugs. Distinguish the inconsistent reports and focus on the consistent ones can save the app developers' time and improve the efficiency. Besides, app developers always do not worry about leaving out bugs reported in inconsistent reports when discarding such reports, because the reports are mostly duplicate \cite{wang2019images} and bugs can always be reported in consistent reports by other experienced crowdworkers. The crowdworkers submitting inconsistent test reports are supposed to be not able to carefully explore the apps under test and find distinct bugs. Further, the large number of test reports are mostly duplicate of the bugs they report, as claimed in many existing studies \cite{gao2019successes} \cite{wang2019images}. Therefore, even if the inconsistent reports really reveal some bugs, such bugs can always be revealed in other test reports. Occasions are quite rare when inconsistent reports can reveal distinct bugs, so eliminating inconsistent reports can save the app developers' time and accelerate the bug reviewing and handling process.}

However, it is challenging to identify inconsistent test reports automatically. Some one-stage approaches \cite{li2020unicoder} \cite{lu2019vilbert} \cite{su2019vl} similarly adopt the encoder-decoder structure, and take the images and texts as input. Such approaches are state-of-the-art tools for image and text consistency detection in general scenarios. While in crowdsourced testing scenarios, such approaches lose effect. First, existing one-stage approaches require large-scale datasets to train the models, and the crowdsourced test reports are hard to reach an adequate volume. Second, for the one-stage approaches, target objects (widgets that indicate or contain bugs) are subtle in the whole app screenshot, and the existing models are always designed to extract the salient object features, and the subtle things would always be neglected by such approaches. However, widgets with bugs on app screenshots are always non-salient and even delicate, making it hardly possible to extract effective features. According to our pilot study, different types of bugs can be revealed from the GUI screenshots, \eg \func, \crash, \etc Different bug types have different visual and textual features, so it is reasonable and more effective to apply different strategies to detect inconsistency for different kinds of bugs.

In this paper, we propose a two-stage approach, \toolname, for crowdsourced test report consistency detection via deep image-and-text fusion understanding. \toolname consists of two stages, the \cs and the \ds, which include \augmentor, \classifier, \decomposer and \detector, respectively.

During the \cs, \toolname first constructs a taxonomy for bugs in the crowdsourced test reports. The taxonomy contains \typeNum types of different bugs (full list in \tabref{tbl:taxonomy}) according to our pilot study, and  the \classifier classifies test reports based on textual descriptions. To construct the taxonomy, we adopt the state-of-the-art pre-trained model for natural language processing (NLP) tasks, BERT \cite{devlin2018bert}. BERT is a pre-trained deep learning model\footnote{\url{https://GitHub.com/google-research/bert}} trained with general scenario datasets. To make the BERT model more suitable for crowdsourced testing, we fine-tune it with a self-constructed large-scale crowdsourced test report dataset and name it as \bert. However, one challenging problem is that the test reports are collected from the real-world industry crowdsourced testing platform, and the data distribution is significantly unbalanced. Therefore, we introduce an \augmentor, which is used to augment the textual descriptions of crowdsourced test reports. To improve the augmenting effect, we design different strategies for textual descriptions of different types of bugs. With the help of the \augmentor, more data can be generated for fine-tuning the \bert model. 

\classifier of \cs consists of the taxonomy and the \bert classification model. As a textual description classifier, we pre-define the bug types. Different from in-house testing, crowdsourced testing is black-box, and most crowdworkers are of different expertise. Crowdsourced testing is conducted from the perspective of app end users, so bugs that can be revealed are limited to several types, \ie \func. Therefore, we conduct an empirical survey to investigate what types of bugs can be revealed in the crowdsourced testing and the frequency. According to the results, we conclude \typeNum types of different bugs, including \func, \crash, \laypro, \dispro, \neterr, \nullsr, \perfp, \errorp, \garerr, and \tranpr. These types cover over 90\% of all test reports. The rest are about user experience, \ie the textual descriptions actually describe the problems in the screenshots, while we do not take them as the ``bugs'' in this paper. Therefore, we hold that the coverage of the constructed taxonomy is wide.

After the bugs are classified by the \classifier, test reports enter the \ds. \decomposer is introduced to analyze the app screenshots and textual descriptions. For app screenshots, \decomposer combines the traditional computer vision (CV) technologies and deep learning models to identify the widgets on app screenshots. Moreover, \decomposer has analysis and understanding of the widgets, including text extraction, and widget type detection. \decomposer also conducts the layout characterization based on the extracted widgets, and the layout characterization presents the relative relationship of all widgets. For textual descriptions, \decomposer first conducts the text segmentation and then performs the dependency parsing on the textual descriptions, extracting the significant words that help locate bugs and corresponding widgets in the app screenshots. We define such words as locating features, including color words ($C$), position words ($P$), text words ($X$), and type words ($Y$) (details in \secref{sec:survey}). By locating features from textual descriptions and extracted widgets with attached information from app screenshots, \detector of \ds can detect the crowdsourced test report consistency.

Bugs revealed from the crowdsourced test reports are of different types, and different types of bugs have different features. For example, the \nullsr problem is always attached with an app screenshot of almost all-black or all-white, the app screenshot of \neterr always contains the texts about network error codes (\eg Error 404), \etc Therefore, we design different strategies for different types of bugs to improve \detector effectiveness. Strategies of \detector are divided into 2 categories, the \textit{General Strategy} and the \textit{Specific Strategy}. For \func, \laypro, \dispro, \tranpr, the \textit{General Strategy} is applied because the screenshots present bugs similarly, and the key is to match the widgets and the text fragments. For \crash, \neterr, \nullsr, \perfp, \errorp, \garerr, each type has very distinct features. Therefore, one \textit{Specific Strategy} is designed for each type of bug. With the designed strategies, the \detector can detect the crowdsourced test report consistency. \vv{The consistency is defined according to the bugs revealed in the app screenshots and presented in the textual descriptions. If one of the bugs revealed in app screenshots can match one of the bugs described in textual descriptions, such a report is considered a consistent one. If none can be matched, the report is considered inconsistent.}

To verify the effectiveness of \toolname, we implement the tool and conduct an empirical experiment with a large-scale crowdsourced test report dataset. The test reports in the dataset are collected from a real-world crowdsourced testing platform and the dataset contains \dataNum test reports. Experiment results show that \toolname can effectively detect consistency in crowdsourced test reports. \vv{Besides, a user study proposed in this paper shows that \toolname is of high practical value in improving the report reviewing efficiency in crowdsourced testing.}

In this paper, we declare to have the following noteworthy contributions.

\begin{itemize}
	\item We propose a novel two-stage approach, \toolname, to detect the crowdsourced test report consistency via deep image-and-text fusion understanding.
	\item We construct a bug taxonomy for crowdsourced test reports based on GUI features. The taxonomy covers the majority of the bugs that can be revealed from crowdsourced testing.
	\item We construct a large-scale crowdsourced test report dataset containing \dataNum test reports.
	\item We implement and release a tool based on \toolname and conduct an experiment to verify the proposed \toolname. The results show that \toolname can effectively detect the consistency of crowdsourced test reports. \vv{A user study further shows the capability for efficiency improvement of \toolname.}
\end{itemize}

\textbf{More information is available on the online package: \underline{\url{https://sites.google.com/view/recode2022}}}.

The rest of the paper is organized as follows. \secref{sec:bg&mv} presents the background information and the motivation of this paper, 
\vv{including an empirical study}. \secref{sec:approach} elaborates the detailed approach design, including four significant components of this paper. In \secref{sec:exp}, the evaluation \vv{and the user study} are illustrated. \secref{sec:rw} presents the related work from two aspects, the crowdsourced testing, and the GUI understanding. In  \secref{sec:con}, the conclusion is made to the whole paper.

\section{Background \& Motivation}
\label{sec:bg&mv}

In this section, we first present the background of this paper, and we then introduce the motivation for this work according to the current situation.

\subsection{Crowdsourced Testing}
\label{sec:ct}

\begin{figure}[!htbp]
    \centering
    \vspace{-0.5cm}
    \includegraphics[width=\linewidth]{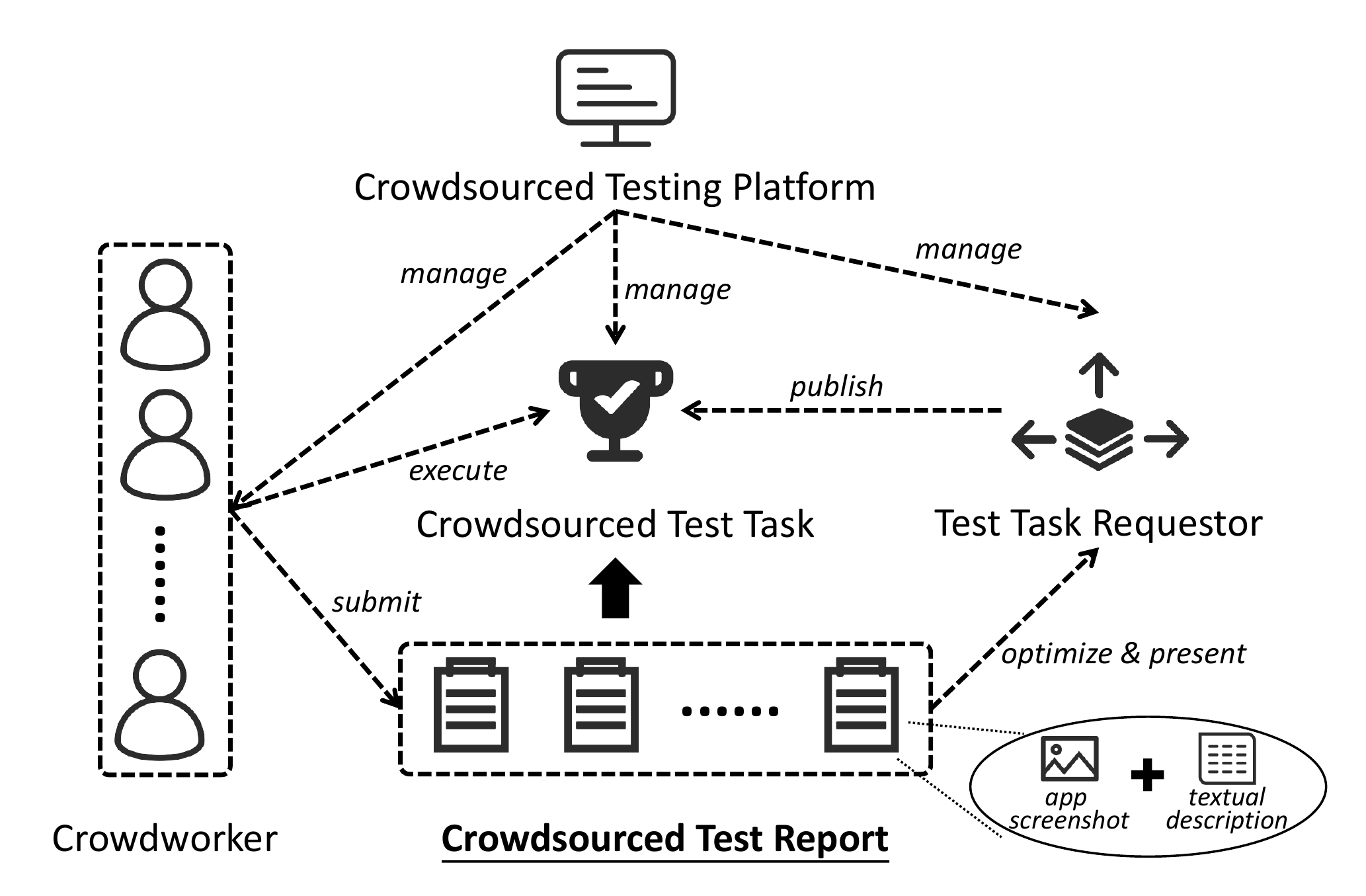}
    \caption{Crowdsourced Testing}
    \label{fig:survey1}
\end{figure}

Crowdsourced testing is a new black-box testing paradigm that recruits crowdworkers to accomplish testing tasks from the perspective of end users \cite{yu2021prioritize}. In most mainstream crowdsourced testing platforms, crowdworkers are required to submit textual descriptions that illustrate the bug behaviors and how they trigger the bugs (operation sequences since the app launch). Besides, they are always encouraged (mostly required) to submit a screenshot that presents the bug occurrence together with the textual descriptions, in order to help app developers review the test reports more intuitively.

In a more common condition, crowdworkers do not have restrictions to be enrolled in the crowdsourced testing work, and meanwhile, they act like common app users, and can freely explore the apps under test without any specific instructions. Therefore some quality control problems of test reports they submit are raised. For example, most of the test reports are duplicate. A more important problem is that the quality of the crowdsourced test reports is widely ranged due to the openness of the crowdworkers. 

\subsection{Inconsistency in Test Reports}

In mobile app crowdsourced testing, quality control of test reports has always been a pain point for improving crowdsourced testing effectiveness. Due to the openness of crowdsourced testing, the involved crowdworkers are with different levels of expertise. Therefore, many \vv{low-quality} test reports are mixed with the \vv{high-quality} test reports and lead to a huge burden for app developers to review the test reports. One of the most significant situations of \vv{low-quality} reports is that the app screenshots and textual descriptions are inconsistent, making the reports completely unusable. \vv{The reason is that app developers have to refer to both app screenshots to identify which app activity has a problem and textual descriptions to learn about how the bug happens. However, app screenshots or textual descriptions cannot singly provide full information for app developers to inspect and reproduce bugs. Therefore, app developers have to combine both pieces of information.}

\begin{figure}[!htbp]
    \centering
    \includegraphics[width=\linewidth]{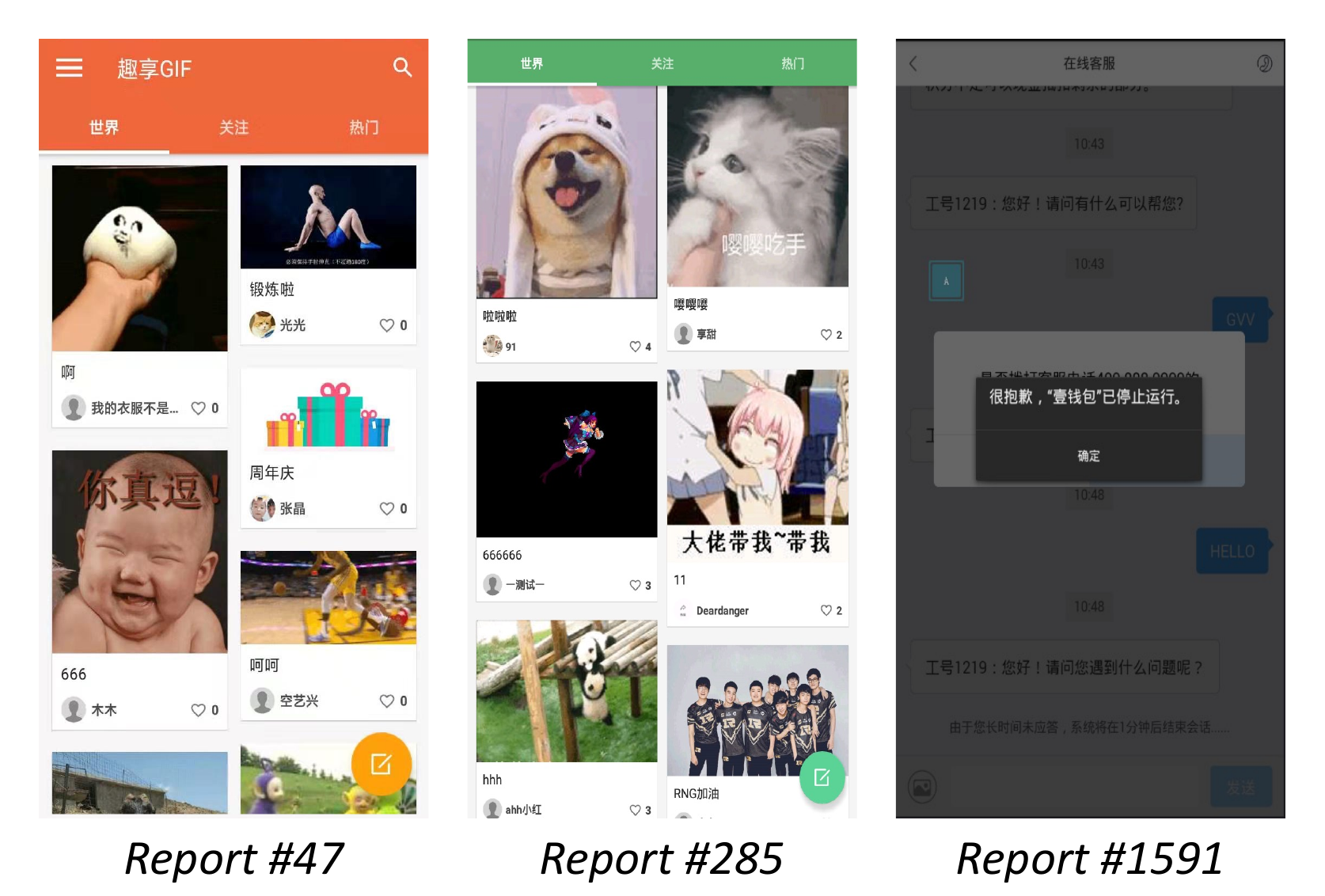}
    \caption{Inconsistent Crowdsourced Test Report Examples}
    \label{fig:example}
\end{figure}

\begin{center}\setlength{\fboxrule}{0.5pt}\fbox{\parbox{0.95\linewidth}{ 
	\ding{202} \emph{Report \#47}: After turning off the ``auto refresh'' button, I cannot manually refresh the contents. \\ 
	\ding{203} \emph{Report \#285}: Randomly click on some widgets in the menu, and return to the ``world'' page, then the app crashes. \\
	\ding{204} \emph{Report \#1591}: Click on the ``telephone'' button at the top-right corner and confirm, the telephone function is not invoked.
}}\end{center}

We give three motivating examples from our large-scale crowdsourced test report dataset (\figref{fig:example}). In the first example (Report \#47), the crowdworker reports a \func, and in the textual description, the crowdworker mentions an ``auto refresh'' button, while there is no such a button (\textbf{no such widget}) on this page as shown in the app screenshot. In the second example (Report \#285), the crowdworker is reporting a \crash, while the attached app screenshot does not show any crash. \vv{The report \#285 is considered inconsistent due to the reason that there is no ``menu'' can be found in the app screenshot. Besides, according to our investigation, this is not the final screen before the app crashes. The crowdworker has to find a replacement screenshot for the crash because such screens can hardly be captured due to the crashes always happen immediately after some actions. App developers can find out how the crashes happen according to the textual descriptions with the aid of app screenshots. In order to determine whether a report regarding the Crash during the dataset labeling, the first criterion is the pop-up windows, which is also the most obvious one. For other reports without pop-up windows in app screenshots, we require the labeling participants to judge carefully whether the app screenshots can match the textual descriptions according to the testing steps provided in the textual description.} In the third example (Report \#1591), the app shows that ``the app has stopped running'', which indicates a \crash, while the textual descriptions are describing irrelevant matter. Therefore, \textbf{the textual description does not contain the bug in the app screenshot.}

Besides the given examples in \figref{fig:example}, there are many other typical kinds of inconsistent crowdsourced test reports, \eg meaningless descriptions like ``1111111...'', which do not indicate any bugs; and complaints about the GUI design, like `` I don't like the color of the button'', which are not the bugs of the apps.

Such inconsistent crowdsourced test reports account for a large proportion of all test reports. We review all the reports in the large-scale crowdsourced test report dataset. Among all \dataNum test reports, only \conNum reports are consistent, which accounts for only \conRate of the dataset (details about the empirical survey in \secref{sec:survey}). Without effective filtering, app developers have to review all the reports. Therefore, we hope to propose an automated tool to help app developers filter out the inconsistent crowdsourced test reports and help them reduce time and resources.

\subsection{Incapability of One-Stage Approach} 

There exist many one-stage deep learning models that are capable of processing image and text data simultaneously \cite{li2020unicoder} \cite{lu2019vilbert} \cite{su2019vl}. However, such models cannot fit into the crowdsourced test report consistency detection task. We conclude with two major reasons.

First, although we have constructed a large-scale dataset, the models still require a much larger scale dataset to reach a relatively good effect. Without a considerable dataset size, the model may not be capable of extracting and learning useful features from the training data. Moreover, to the best of our knowledge, the dataset we contribute in this paper is the largest crowdsourced test report dataset with labeled information both in the academic community and industry. While with such a dataset, the model still cannot effectively detect the inconsistency of crowdsourced test reports according to our practice.

Second, for most crowdsourced test reports, the bugs shown on app screenshots will not occupy a large proportion of the whole page. For example, if a \crash appears, the prompt will only show on the center of the page; a \func will only link to a single \texttt{Button} or \texttt{TextView}. Therefore, during the encoding or embedding of the app screenshots, areas indicating bugs would be ignored, and corresponding features will not be extracted and learned.

Due to the aforementioned incapability of one-stage approaches, we have come up with a two-stage approach. The approach classifies the bugs based on the bug features and detects crowdsourced test report consistency with different strategies and deep GUI image understanding, which can be more effective.

\subsection{\vv{Empirical Survey}}
\label{sec:survey}

For detecting strategies of \detector, the premise is that bugs can be revealed from app screenshots, and textual descriptions can be analyzed to locate the bugs and corresponding buggy widgets. Therefore, we conduct an empirical survey on the mobile app crowdsourced test reports. The data collection process is elaborated in \secref{sec:data}.

We first investigate the consistent situation of all crowdsourced test reports, which is the basis of this work's motivation. Of the \dataNum crowdsourced test reports, only \conNum reports are consistent. In other words, the textual descriptions of the reports actually describe the bugs revealed in the app screenshots submitted by crowdworkers. The fact that \textbf{only \conRate of the reports are consistent} reflects the severe quality problem in crowdsourced testing. The large percentage of inconsistent test reports can lead to a waste of time for app developers in reviewing the test reports. 

\begin{figure}[!htbp]
    \centering
    \vspace{-0.5cm}
    \includegraphics[width=\linewidth]{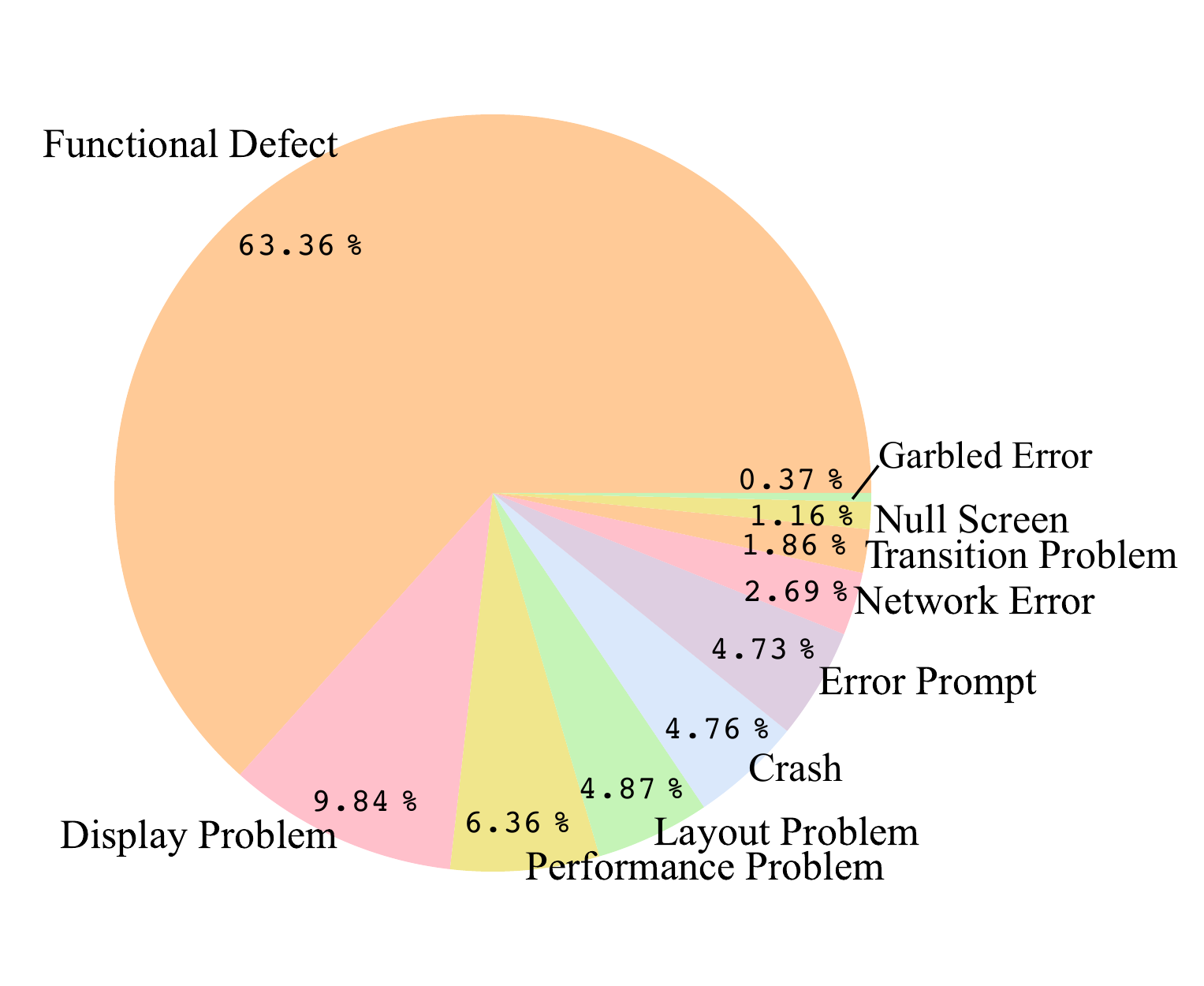}
    \vspace{-0.5cm}
    \caption{Bug Distribution among Mobile App Crowdsourced Test Reports}
    \label{fig:survey1}
\end{figure}

Also, we conclude some common inconsistent types. First, crowdworkers submit meaningless test reports, which have nothing to do with the bugs, because some tasks reward them only according to the report number they submit. Second, the crowdworkers report issues about GUI designs, which may affect user experience, while for app developers, such issues are not considered as bugs. Third, some crowdworkers cannot precisely describe the problems in the app screenshots due to a lack of experience or expertise, making the submitted reports hard to understand for app developers. 

The second target of the empirical survey is to investigate the bug types revealed in crowdsourced testing. \vv{This target is investigated on the \conNum consistent crowdsourced test reports.} \vv{Our dataset is one of the largest datasets in crowdsourced testing research, and it is representative.} According to the results (\figref{fig:survey1}), the most common bug type in crowdsourced testing is \func, which accounts for 63.36\%. The following bug type is \dispro, accounting for 9.84\%, and \perfp, accounting for 6.36\%. The least common bug type is \garerr, which only makes up about 0.37\% of the test reports. Crowdworkers always report \func, which relates to the app business logic, much more frequently, because \func most directly affects their user experience. Besides, \crash, \laypro, \dispro, \perfp, and \errorp also have negative effects to end users, while such problems often happen due to the compatibility problems introduced by the well-known ``fragmentation problem'' \cite{yu2021layout} \cite{wei2016taming} on mobile platforms, so only part of crowdworkers would report such problems. \vv{For different bug categories, there are different features. For categories with distinct features, we adopt \textit{Specific Strategies} for each category. Some categories do not have distinct features, so we adopt the \textit{General Strategy} for them. Such categories cannot be merged because the taxonomy is based on the app behaviors and different bug categories have different textual features. Besides, some similar types, like the Layout Problem and the Display Problem, are not essentially the same. The Layout Problem refers to the layout problems of existing widgets, which are normally presented, while the Display Problem refers to the situation in which some widgets cannot be normally presented, including missing or overlapping.}

However, some problems cannot be revealed in crowdsourced testing. For example, the \texttt{NullPointerException} may not be directly exposed to end users and will be presented in another form from the app GUI. The exceptions will only be recorded in logs.

\begin{figure}[!htbp]
    \centering
    \includegraphics[width=\linewidth]{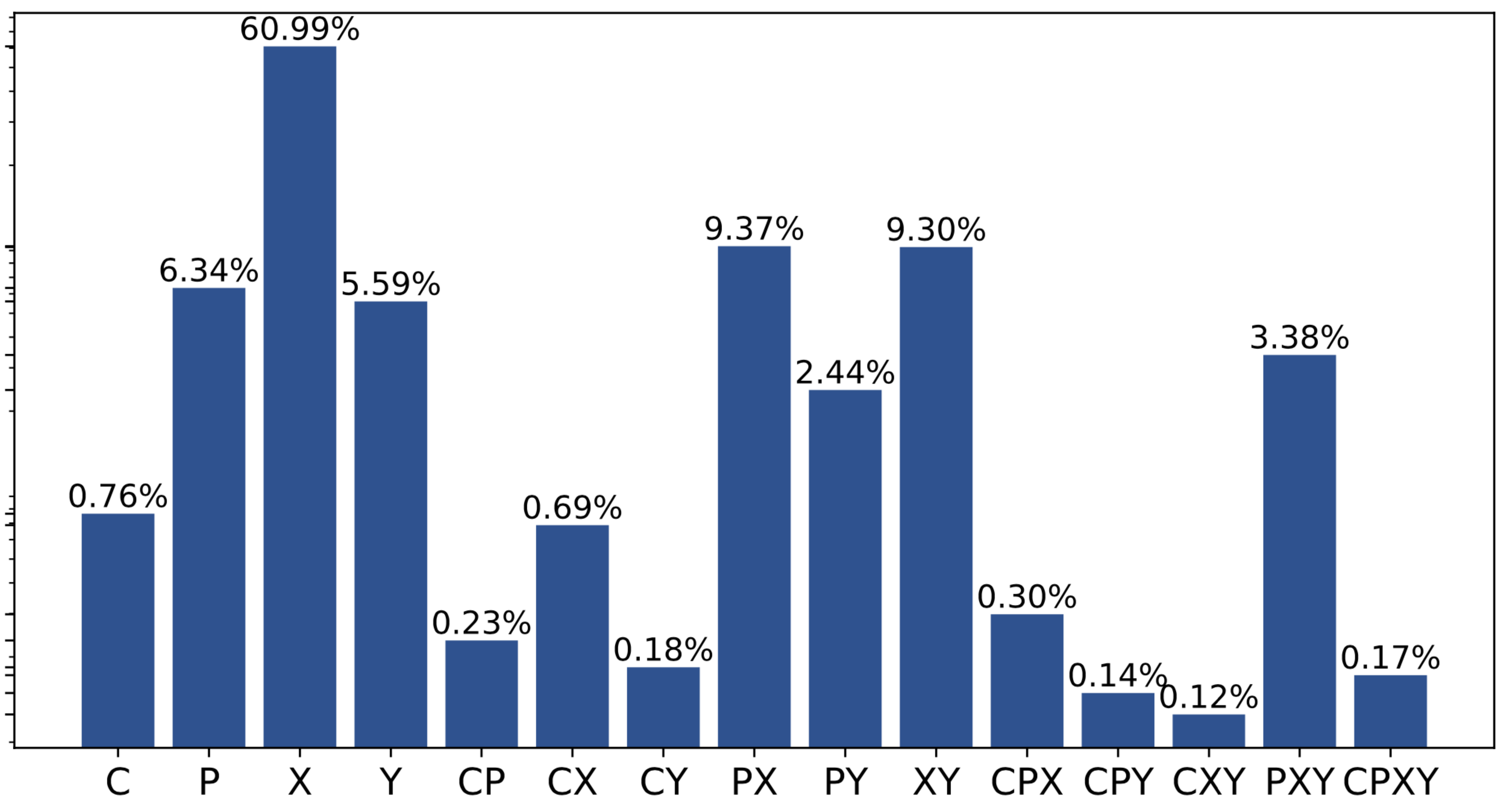}
    \caption{Locating Features in Textual Descriptions\protect\footnotemark}
    \label{fig:survey2}
\end{figure}
\footnotetext{The number of letters indicates the types of used locating features. C - color words; P - position words; X - text words; Y - type words.}

Another problem we focus on is how crowdworkers locate the target widgets with bugs in textual descriptions. This also determines how we build the locating feature dataset in the \detector. 

According to the manual labeling results, we find that most crowdworkers use four different categories of locating features, the color words ($C$), the position words ($P$), the text words ($X$), and the type words ($Y$). Text words are most intuitive and thus most widely used by crowdworkers, and 84.31\% of the crowdsourced test reports use or partly use text words as locating features. 2.59\%, 22.21\%, and 21.31\% of the test reports use or partly use color words, position words, and type words, respectively. Moreover, the three locating features are used together with text words to help locate the buggy widgets more accurately.

As to the locating feature number in the test reports, we find that most crowdworkers locate the buggy widgets with one locating feature, which accounts for 73.68\% of all crowdsourced test reports. Crowdworkers are unprofessional end users of mobile apps. Therefore, it is reasonable that they use single locating features to describe the widgets with bugs. However, there are still 0.17\% test reports that use four locating features, which indicates some crowdworkers are with rich experience and high expertise. Interestingly, all reports containing four locating features in textual descriptions are consistent. Besides, 22.21\% and 3.94\% test reports contain two and three locating features in textual descriptions, respectively. Such data shows that using one or two locating features (mostly with text words) is within the majority crowdworkers' capability. Some experienced crowdworkers can provide valuable consistent test reports with detailed textual descriptions.

\section{Approach}
\label{sec:approach}

\begin{figure*}[!htbp]
    \centering
    \includegraphics[width=\linewidth]{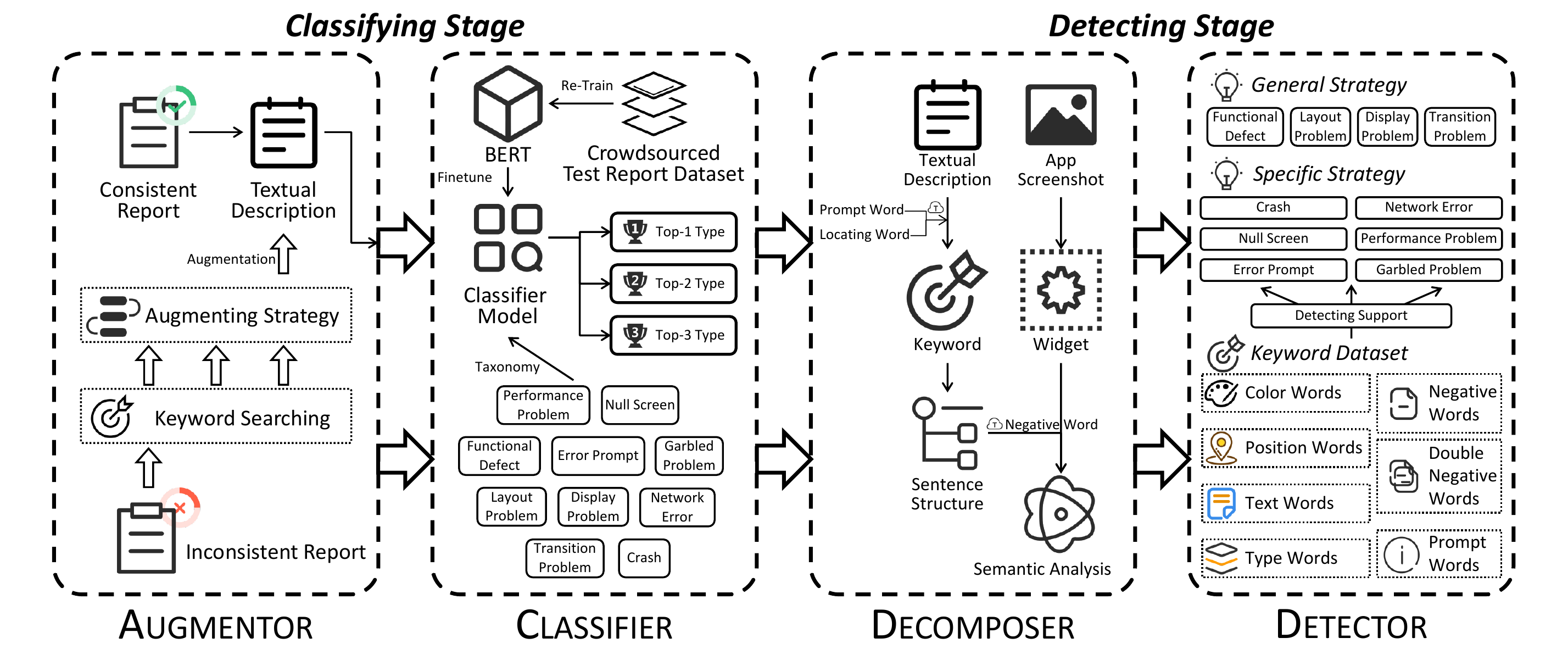}
    \caption{\toolname Framework}
    \label{fig:framework}
\end{figure*}

This section presents the design details of \toolname, which is used for mobile app crowdsourced test \underline{\sc \textbf{Re}}port \underline{\sc \textbf{Co}}nsistency \underline{\sc \textbf{De}}tection via deep image-and-text fusion understanding. \toolname is a two-stage approach and is composed of \cs and \ds. The \cs includes an \augmentor and a \classifier, and the \ds includes a \decomposer and a \detector (\figref{fig:framework}). For a given crowdsourced test report, \toolname first classifies the report into a specific bug type with the \classifier according to the textual descriptions and then utilizes the \decomposer to extract the locating features. Finally, the \detector detects the consistency of crowdsourced test reports based on the strategies. \vv{In order to judge whether a report is consistent or inconsistent, we refer to the following equation: for a crowdsourced test report, suppose the set containing bugs revealed in the app screenshot is $A$, and the set containing bugs described in the textual description is $T$. If $A \cap T \neq \emptyset$, the crowdsourced test report is consistent, otherwise it is inconsistent.}

\subsection{\toolname Augmentor}
\label{sec:augmentor}

\begin{table}[!htbp]
	\centering
	\caption{Bug Type in \toolname}
	\scalebox{1}{
	\begin{tabular}{c|c|c}
	\toprule
	\textbf{Type} & \textbf{\# Initial Report\tablefootnote{number before augmentation}} & \textbf{\# Final Report\tablefootnote{number after augmentation, used in taxonomy construction}} \\ \midrule
	Functional Defect   & 2,526 & 480\tablefootnote{the data is enough so we do not conduct the augmentation but conduct a random selection for \classifier training} \\
	Crash               & 205   & 497   \\
	Layout Problem      & 210   & 301   \\
	Display Problem     & 424   & 424   \\
	Network Error       & 116   & 481   \\
	Null Screen         & 50    & 294   \\
	Performance Problem & 274   & 454   \\
	Error Prompt        & 204   & 367   \\
	Garbled Error       & 16    & 72    \\
	Transition Problem  & 80    & 370   \\  \midrule
	Total               & 4,105 & 3,740 \\ \bottomrule
	\end{tabular}}
	\label{tbl:taxonomy}
\end{table}

To train the \classifier of \toolname, we label the consistency or inconsistency of the crowdsourced test reports (labeling details in \secref{sec:data}). We find that only \conNum test reports are consistent, and the data distribution of different types of bugs is uneven. Therefore, the data of some types can be inadequate to train the deep learning model-based \classifier, so the data of some types should be augmented to reach an even distribution\vv{, with the aim of avoiding any negative effects brought by the imbalanced data distribution. Many studies \cite{chawla2002smote} \cite{han2005borderline} \cite{shen2016relay} have pointed out that imbalanced data distribution is a severe problem in deep learning approaches. The performance of prediction in categories with fewer samples will be much worse than that of categories in more samples. Data distribution re-balancing is a common practice to solve the problem.}

To reach an even distribution and improve the model effectiveness, we propose an \augmentor, which is used to augment the textual descriptions. According to our review of the crowdsourced test reports, we use the keyword replacement method to augment the textual descriptions for different types of bugs. The resources for \augmentor are GitHub issues and fixing commits, app reviews from app stores (\ie Google Play), test reports from the industry, and Stack Overflow discussions. \vv{Specific keywords are extracted from the external resources to replace the synonyms in the existing reports. After the replacement, a manual check is conducted on each augmented textual description to verify the correctness and whether to keep such augmented textual descriptions.}

\vv{In order to collect the required information, we first refer to the AndroZooOpen \cite{liu2020androzooopen} dataset, which is a very important research dataset containing over 45,000 open-sourced Android app repositories. We then select the repositories whose star number is higher than 500 as our investigation objects, which number is 1179. With the apps selected, we further find the corresponding resources of such apps, like GitHub issues and fixing commits, app reviews from app stores, and Stack Overflow discussions. Other studies are also referred to, including \cite{wendland2021andror2} \cite{allix2016androzoo} \cite{liu2020androzooopen} \cite{scalabrino2019data} \cite{geiger2018graph} \cite{wang2018android}.}

For \func textual descriptions, the number is much larger than other types of bugs, so we do not augment the \func textual descriptions. During the \classifier construction, we randomly select approximately the same number of \func textual descriptions as other types of bugs. \vv{The process to the reports in Functional Defect is an under-sampling process with randomness, which may bring data bias. For the reports of Functional Defect, the textual descriptions are of similar structures, including bug behaviors and expectations of app behaviors. The features of the textual descriptions in crowdsourced test reports eliminate such possible bias. }
For the rest categories, we adopt the following strategies.

\begin{itemize}
	\item For \crash textual descriptions, we search for the keywords indicating crashes, like ``abnormal exit'', and ``app crash'', from the \augmentor resources to complete the augmentation.
	\item For \laypro and \dispro textual descriptions, we extract the keywords, such as ``display'', ``show'', and ``present'', to pick textual descriptions from \augmentor resources to augment the \laypro and \dispro textual descriptions.
	\item For \neterr textual descriptions, we focus on the keywords like ``network'', and the HTTP error code like ``404 not found'', ``502 bad gateway''. Such keywords can help us find more textual descriptions from the \augmentor resources to augment the \neterr type textual descriptions.
	\item For \nullsr textual descriptions, the keywords include ``white screen'', ``null screen'', ``black screen'' are used for collecting the textual descriptions of \nullsr from \augmentor resources.
	\item For \perfp textual descriptions, we search for the textual descriptions of \augmentor resources with the keywords like ``stuck system'', ``long time'', ``slow'' and so on.
	\item For \perfp textual descriptions, we search for the textual descriptions of \augmentor resources with the keywords like ``stuck system'', ``long time'', ``slow'' and so on.
	\item For \errorp textual descriptions, they usually appear with a pop-up window telling the error information. The crowdworkers tend to report what is showing in the pop-up window. Therefore, we collect \augmentor resources to augment the dataset with keywords like ``prompt'', ``show'', ``pop-up'' \etc 
	\item For \garerr textual descriptions, we first find textual descriptions containing the ``garbled texts'' and then translate the original textual descriptions to another language and translate back with machine learning to have different descriptions.
	\item For \tranpr textual descriptions, we search with keywords ``transit'', ``back'', ``exit'', in the \augmentor resources and augment the dataset. 
\end{itemize}

\vv{Potential noise data might be brought by the augmentation with the automated approach. In order to eliminate such potential, we conduct a manual check. During the augmentation, we refer to some external resources, including GitHub issues and fixing commits, app reviews from app stores, test reports from the industry, Stack Overflow discussions, and datasets in previous studies. Such a practice can improve the reliability of the augmentation. The resources are utilized to conduct the keyword replacement approach as the \augmentor. After the augmentation, we manually review all the generated textual descriptions and filter out the ones that cannot reflect bugs.}

With the proposed augmenting strategies, we build the \augmentor of \toolname and augment the textual descriptions of each type of bug to a relatively equal number. The number of each type of bug is shown in \tabref{tbl:taxonomy}. 

\begin{table*}[!htbp]
	\centering
	\caption{Detailed Taxonomy Adopted in \toolname}
\scalebox{1}{
\begin{tabular}{c|c|c|c}
\toprule
\textbf{Bug Type}            & \textbf{Example (Simplified)}                                                                  & \textbf{Feature}                                                          & \textbf{\%} \\ \midrule
Functional Defect   & \tabincell{c}{``The ID number textbook do not \\ have any restrictions.''}                                & Describing the app functions                                     & 63.36\%     \\ \midrule
Crash               & ``The app crashes.''                                                                    & App crashes or exit abnormally                                   & 4.76\%     \\ \midrule
Layout Problem      & ``The button should stay in the former button''                                         & Widget layout is different from expectations                     & 4.87\%     \\ \midrule
Display Problem     & ``The texts are overlapped each other.''                                                & Widgets (texts, images) are overlapped                           & 9.84\%     \\ \midrule
Network Error       & \tabincell{c}{``I cannot connect to server \\ when I have good network link.''}                           & Cannot load resourced from out domains                           & 2.69\%     \\ \midrule
Null Screen         & ``The app screen is all-white''                                                         & The screen shows nothing                                         & 1.16\%     \\ \midrule
Performance Problem & \tabincell{c}{``I wait for a long time \\ and cannot have the page loaded.''}                             & The response time is much longer than expected                   & 6.36\%     \\ \midrule
Error Prompt        & ``The app indicates the ‘SQL ERROR’.''                                                  & There is explicit prompts indicating the errors                  & 4.73\%     \\ \midrule
Garbled Error       & ``There are all garbled texts.''                                                        & Caused by the mismatched character sets                          & 0.37\%     \\ \midrule
Transition Problem  & \tabincell{c}{``The app should show the personal information \\ page while it shows the ‘Moment’ page.''} & \tabincell{c}{App transitions among activities \\ are different from expectations} & 1.86\%     \\ \bottomrule
\end{tabular}}
\label{tbl:detail}
\end{table*}

\subsection{\toolname Classifier}
\label{sec:classifier} 

To construct the \classifier, it is necessary to first build a taxonomy. We investigate the crowdsourced test report dataset with \dataNum reports, and we conclude in total \typeNum types of bugs that will be revealed in crowdsourced testing from the perspective of crowdworkers as end users. We present the bug types, simplified examples, features, and the accounting percentages in \tabref{tbl:detail}. The bugs are classified on the basis of the bug features reflected in GUI and textual descriptions.

\begin{itemize}
	\item \textbf{\func} is closely related to the business functions of the apps, and the apps behave differently from the expectations of the users. Some are business logic errors, and the app runs normally. Some are programming bugs, and the apps may throw exceptions.
	\item \textbf{\crash} has a significantly negative effect on user experience and is one of the most severe types of bugs. Many reasons, including memory leaks, and hardware compatibility, will lead to app crashes.
	\item \textbf{\laypro} will happen due to hardware incompatibility. Some widgets will not show as expected. For example, the text overlap, widget occlusion, blurred screen \cite{liu2020owl}.
	\item \textbf{\dispro} refers to the problem of missing widget contents, like images, or even missing widgets. Such problems will make users not able to go on with using the apps.
	\item \textbf{\neterr} is related to the network condition. Typical examples include different kinds of HTTP errors. Some are due to the client having a weak network link, and others are because the servers have errors and cannot provide service. 
	\item \textbf{\nullsr} is an obvious type of bugs. A large proportion of the app activity is all-black or all-white.
	\item \textbf{\perfp} means that the requests from app users will not receive responses in time. The apps keep loading the resources or have the users wait for a long time.
	\item \textbf{\errorp} always comes with pop-up windows, and the messages about the bugs will be presented to app users.
	\item \textbf{\garerr} is caused by the encoding and decoding of different character sets. Such bugs make app users cannot understand the textual information from the apps.
	\item \textbf{\tranpr} means the transitions among different app activities are not meeting the expected destination activity, and such problems will greatly confuse the app users.
\end{itemize}

\begin{figure}[!htbp]
    \centering
    \includegraphics[width=\linewidth]{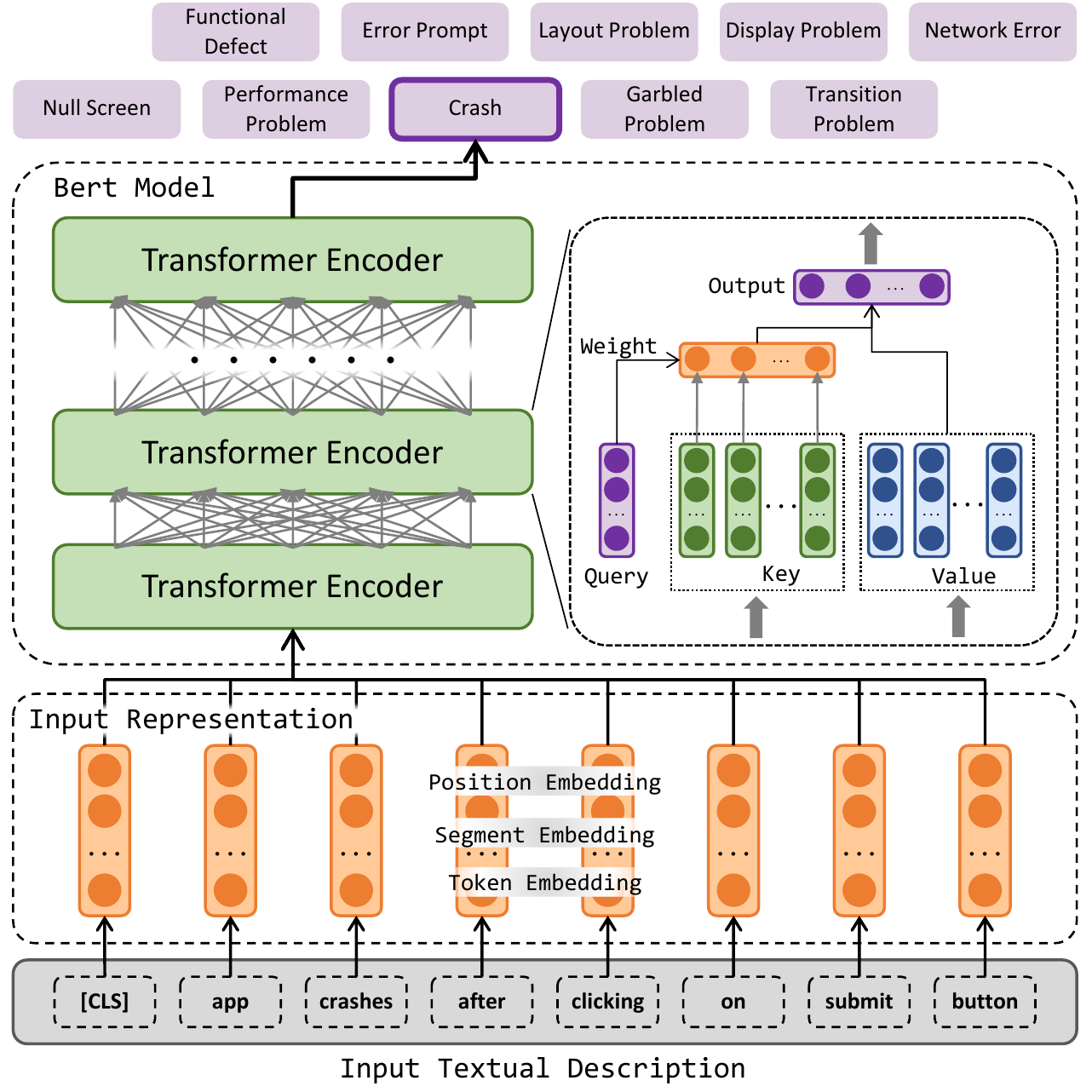}
    \caption{\classifier Structure (\bert)}
    \label{fig:classifier}
\end{figure}

In total, we conclude \typeNum different bug types that are commonly proposed in crowdsourced testing. As illustrated in \secref{sec:ct}, crowdsourced testing is a user-end black-box testing paradigm, so specific bug types can be reported. According to our survey (\secref{sec:survey}), the \typeNum bug types contained in our taxonomy would cover the vast majority of the bugs in crowdsourced testing.

With the pre-defined taxonomy, we build the \classifier on the basis of the BERT model \cite{devlin2018bert}. BERT is a pre-trained model, and we fine-tune the model with the augmented data and rename the model as \bert. 
\vv{The \bert model structure is from the BERT model, which utilizes several encoders of the Transformer model. Each encoder of the Transformer model, implementing with the attention mechanism. The attention mechanism calculates the representation of each textual sentence by considering the position relationship among the sentences. The attention mechanism relies on three main vectors, query Q, key K, and value V, by mapping a query and a set of key-value pairs to an output vector. We employ a scaled dot-product self-attention to calculate the attention scores of each token by taking the dot product between all of the query vectors and key vectors. The attention scores are then normalized to probabilities using the SoftMax function to get the attention weights. Finally, the value vectors can be updated by taking a dot product between the value vectors and the attention weight vectors. The self-attention operation is computed using three matrices Q, K and V as follows: $Attention(Q, K, V)\ = softmax(\frac{{\rm QK}^T}{\sqrt{d_k}}) V$. To capture richer semantic meanings of the input textual descriptions, we use a multi-head mechanism to implement the self-attention, which allows the model to jointly attend the information from different sentence representation subspaces at different positions. For d-dimension Q, K, and V, those vectors are split into h heads where each head has d/h-dimension. After all of the self-attention operations, each head will then be concatenated to feed into a fully-connected layer including two linear transformations.}
The detailed model structure of the \classifier is shown in \figref{fig:classifier}. For the textual descriptions, we first conduct the word embedding, including Token Embedding, Segment Embedding, and Position Embedding. \vv{The input of the model is represented with the sum of three embeddings.} 
Token Embedding \vv{is the embedding vectors of the words themselves, and} transfers each word in the textual descriptions into fixed-length vectors (768 in \bert). 
Segment Embedding processes the sentence pairs to determine whether they are supposed to be classified into the same category. \vv{It embeds the paired sentences, which is the concatenation of two vectors representing two sentences, which use all ``0''s to indicate the words in the first sentence and all ``1''s to indicate the words in the second sentence. For the classification tasks with the BERT model, the sentences are paired as inputs of the model to identify whether they belong to the same category.}
Position Embedding \vv{refers to encoding the location information of all words, and it} gives vectors representing the position of each word in the sentence and helps the model obtain information from the context. 
Then the encoded vectors are fed into the Transformer \cite{vaswani2017attention} encoders connected in sequence. The encoders adopt the attention mechanism \cite{mnih2014recurrent}. During the model training, we follow the common practice and divide the data at the ratio of 6:2:2 for the training set, validation set, and test set. We set the batch size as 32, and the learning rate is $5 \times 10^{-5}$. The total training epoch is 30. \vv{All the hyper-parameters are the same as the original BERT model presented in \cite{devlin2018bert}.}

For a crowdsourced test report, the reported bug may be classified into different types \vv{because the textual descriptions may contain more than one bug reported by the crowdworker}. For example, in some app designs, the \crash bug will be accompanied by the \errorp, saying ``the app crashes'' in a pop-up window. Therefore, it is not sound to simply classify the test report into one type. \vv{During the data labeling, we confirm one bug by referring to both app screenshots and textual descriptions in the report. However, considering the fact that more than one bug might be reported in the textual descriptions,} we collect the top-3 results with the highest confidence \vv{from the \classifier prediction} to eliminate the negative effect from the ambiguity of crowdsourced test reports.

\subsection{\toolname Decomposer}

\decomposer is designed to extract the key semantic information from both app screenshots and textual descriptions.
 
App screenshots can be seen as a set of widgets from the perspective of app users. Therefore, the first processing step on app screenshots is to extract the widgets. In \toolname, we construct a widget-extracting model utilizing the advantages of traditional CV technologies and deep learning models. Widgets on app screenshots are divided into text widgets and non-text widgets. Text widgets are extracted by OCR algorithms and are linked to non-text widgets. For example, the ``submit'' text on the button. Text widgets are significant to directly conduct matching with the textual descriptions from crowdsourced test reports. However, some widgets are not accompanied by text. We feed such widgets into a DL model proposed by Yu \etal \cite{yu2021prioritize}, which can identify the type of the widgets.
\vv{Specifically, the model is modified from the classic VGG-16 deep learning model. Specifically, the proposed model can identify the Android GUI widgets of 14 different types, which are the most widely used ones, including Button, CheckBox, CheckTextView, EditText, ImageButton, ImageView, ProgressBarHorizontal, ProgressBarVertical, RadioButton, RatingBar, SeekBar, Switch, Spinner, and TextView. The model is trained with a dataset containing 36,573 widget screenshots. The ratio of the training set, validation set, and test set is 7:1:2, which is a common practice for an image classification task. The neural network is composed of multiple Convolutional layers, MaxPooling layers, and FullyConnected layers. AdaDelta algorithm is used as the optimizer, and this model adopts the categorical\_crossentropy loss function. For the 14 categories on the test set, the overall accuracy reaches 89.98\%. For each specific widget category, the average precision is 90.05\% (ranging from 74.36\% to 99.81\% among 14 categories), the average precision is 74.36\% (ranging from 70.83\% to 100.00\% among 14 categories), and the average F1 score is 89.92\%. We do not calculate the metrics of the model because the datasets used in previous work and in \toolname are highly similarly featured and distributed, and some other studies like \cite{yu2021prioritize} and \cite{yu2019crowdsourced} also prove the generalizability of the model.}
Further, in the textual description, crowdworkers will use the position relationship of other widgets to describe the problem widget. Therefore, we characterize the app screenshot layout based on the extracted widgets.

Textual descriptions contain the locating features that help app developers locate the position of the problem widgets in the app screenshots. \decomposer is used to extract such locating features from the textual descriptions. Locating features include color words ($C$), which describe the widget colors; position words ($P$), which describe the widget positions on the app screenshots; text words ($X$), which describe the texts on the target widgets; and type words ($Y$), which describe the widget types, \ie \texttt{Button}, \texttt{TextField}. We build the datasets for the locating features, respectively, with reference to different data sources, including GitHub issues and fixing commits, app reviews from app stores, and test reports from the industry. 

Besides, we build the emotional identifying dataset, which stores negative words and double-negative words. \vv{Negative words are collected from existing studies \cite{hu2004mining} \cite{liu2005opinion}, including ``should not exist'', ``abnormal'', ``non'', ``no'', ``disappear'', ``miss'', ``lose effect'', ``unverified'', ``different'', ``hardly'', ``lack'', ``cannot'', and ``dislike'', \etc Double-negative words are always combinations of even numbers of negative words. We also refer to GitHub issues and fixing commits, app reviews from app stores, and test reports from the industry to complement the dataset with negative words that are frequently used in the software engineering field.} We also build the prompt word dataset, containing words that indicate the existing texts on app screenshots that will appear in the textual descriptions. \vv{Examples are like ``indicate'', ``say'', ``present'', ``show'', ``prompt'', ``hint'', ``reveal'', ``demonstrate'', and ``inform''.} Prompt words can help link app screenshots with textual descriptions.

With the dataset of locating features, negative words, and prompt words, we analyze the basic sentence relationships of the textual descriptions. We adopt the open-sourced dependency parsing analysis tool, DDParser, from Baidu\footnote{\url{https://GitHub.com/baidu/DDParser}} \cite{dozat2016deep} \cite{zhang2020practical}. Dependency parsing analysis \cite{zhu2013fast} aims at analyzing the dependency relationship of the words to determine the sentence structure. For textual descriptions of crowdsourced test reports, the widely used sentence structures include \cite{zhang2020practical}: \underline{\textbf{SBV}} (Subject-Verb structure), \underline{\textbf{VOB}} (Verb-Object structure), \underline{\textbf{ADV}} (Adverbial structure), \underline{\textbf{CMP}} (Complement structure), \underline{\textbf{ATT}} (Attributive structure), and \underline{\textbf{F}} (Position-Word structure). By dependency parsing analysis, each word is labeled with the sentence composition. Afterward, we extract the keywords from the textual descriptions, and identify the prompt words, locating features and negative words according to the constructed dataset. If the sentence contains prompt words, we identify the prompt texts from the textual descriptions with the sentence structure; if the sentence does not contain prompt words, we observe that crowdworkers tend to describe the widgets with bugs, and we extract the key nouns and the corresponding qualifiers. 

Then, with the extracted words from textual descriptions, we identify the emotional tendency of the textual descriptions based on the negative words and double-negative words. \decomposer decomposes the app screenshots and textual descriptions into widgets and keywords, respectively. Of the keywords from textual descriptions, we extract prompt words, which indicate the existing texts on app screenshots, locating features, which describe the features of the mentioned widgets on app screenshots, and (double-)negative words, which show the emotional tendencies of the textual descriptions. Acquiring such information can lay a solid foundation for the \detector to detect the crowdsourced test report consistency.

\subsection{\toolname Detector}

Different types of bugs have significantly different app screenshot features. Therefore, \toolname is designed to adopt a two-stage approach. \detector works on the basis of results from \classifier and \decomposer. Different detecting strategies are designed to detect the mobile app crowdsourced test report consistency. Such strategies are divided into two categories, the \textit{General Strategy} and the \textit{Specific Strategy}. The \textit{General Strategy} is designed for \func, \laypro, \dispro, and \tranpr. The four types of bugs do not have distinct features on app screenshots. For the rest types of bugs, each type has distinct features revealed from app screenshots, so we design the \textit{Specific Strategies} for each type in \detector. With the estimation of the strategies, a score denoted as $S_{dt}$, is assigned to each test report.

The \textit{General Strategy} is designed for \textbf{\func}, \textbf{\laypro}, \textbf{\dispro}, and \textbf{\tranpr}. Texts are most widely used to conduct the match, while not all widgets can be identified by existing texts, and some widgets are not attached with texts, or more than one widget shares the same texts. Therefore, we introduce the locating features, which contain more features to identify the widgets. Locating features include color words, position words, type words, and text words. Color words indicate the colors of the target widgets; position words describe the approximate positions, like the top-left corner of the screenshot; and type word means the widget type, \ie \texttt{Button} or \texttt{TextView}. The locating features can be combined to identify the widgets. For example, ``the widget on the \underline{left} of the \underline{green} \underline{`confirm'} \underline{button}...'', the textual description contains position words, color words, text words and type words in turn. For each widget extracted from textual descriptions by \decomposer, \detector identifies whether the four locating features exist. If so, widgets from app screenshots are matched by each locating feature. Each locating feature is assigned to a weight, $\omega_C$, $\omega_P$, $\omega_X$, $\omega_Y$\footnote{$\omega_C + \omega_P + \omega_X + \omega_Y = 1$}, and if any locating feature is matched, the corresponding weight is counted, otherwise truncated. The sum of four locating feature weights is the widget score, and the $S_{dt}$ is assigned to the average of all the widgets extracted from textual descriptions.

For the rest types of bugs, we design \textit{Specific Strategies} for each type based on the bug features revealed in app screenshots.

\vv{\underline{For \textbf{\crash}}}, app developers always design a pop-up window that indicates the app crashes. The pop-up contains texts like ``no response'', ``stop running'', \etc Therefore, we first extract the pop-up windows from the app screenshots. Then, OCR technologies are adopted to extract the texts from the pop-up windows. If such texts contain the corresponding words related to crashes, \detector assigns the $S_{dt}$ to 1, otherwise 0.

\vv{\underline{For \textbf{\neterr}}}, situations are divided into 2 categories. First, the embedded H5 pages may encounter the \neterr, and the app screenshots contain the error HTTP response status codes, like 404, 502 \etc We extract texts from app screenshots and locate the error HTTP response status codes to determine whether the app screenshots reflect a \neterr. Second, for native app activities, \neterr is shown by pop-up windows. We also use OCR technologies and match with keywords like ``server error'', ``unable to link'' \etc Such words indicate the \neterr existence. If the keywords extracted from app screenshots \textbf{contains} texts from textual descriptions, \detector assigns the $S_{dt}$ to 1.

\vv{\underline{For \textbf{\nullsr}}}, the feature is obvious. The whole app activity is all-white or all-black. According to this feature, we binarize the app screenshots into black-and-white ones. The white pixels refer to the widget borders, and the black pixels refer to the backgrounds. Then, we detect the continuous areas without widget borders. If there exist one or more areas that account for over a certain ratio of the whole app screenshot in size, the $S_{dt}$ is assigned the value 1; otherwise 0. The certain ratio is represented by $\theta$, and we set it as 0.75 in our \toolname implementation.

\vv{\underline{For \textbf{\perfp}}}, app clients cannot receive expected responses from the servers, and the contents are waiting to be loaded. Therefore, the loading icons or texts are the distinct features of \perfp. Loading icons are of similar shape features, so we collect different loading icon styles from different apps and construct a loading icon dataset. For app screenshots in crowdsourced test reports, we extract the widgets and match them with the dataset. If one or more widgets on app screenshots are matched, we confirm the existence of \perfp. Moreover, some apps use texts to directly indicate the \perfp, including ``loading'', ``load failed'', \etc Existence of such words lets \detector assign the $S_{dt}$ to 1, otherwise 0.

\vv{\underline{For \textbf{\errorp}}}, there is a high probability that texts exist to indicate system errors. Also, crowdworkers would describe the corresponding prompt texts in textual descriptions. Therefore, we extract the error prompts, which always follow the prompt words and are presented as subordinate clauses. The detection of prompt words is on the basis of the prompt word dataset. For app screenshots, we adopt OCR technologies to extract texts. If texts from app screenshots \textbf{contain} error prompts extracted from textual descriptions, the $S_{dt}$ is assigned the value 1, otherwise 0.

\vv{\underline{The \textbf{\garerr}}} are always triggered by the wrong character encoding and decoding with different source and target sets. Therefore, we first extract all the texts from the app screenshots \vv{and identify whether each character} belongs to the normal characters (UTF-8 set). If there exists any character that does not belong to the preset character sets, the $S_{dt}$ is assigned the value 1, otherwise 0.

\vv{In order to design reasonable and generalizable features as detecting strategies in the scenario of crowdsourced testing, we refer to different resources. The first data source is the crowdsourced test reports to identify the information about the above features. To further enhance the generalizability of the proposed features, we refer to other information, including GitHub issues and fixing commits, app reviews from app stores, test reports from the industry, and Stack Overflow discussions.} 
\vvvvv{Besides referring to different resources, we have to consider the special condition of reports in the crowdsourced testing scenario. In crowdsourced testing, reports are submitted by crowdworkers, who are more like common app users instead of professional testers. Because crowdworkers are not professional testers, they prefer submitting app screenshots with obvious indicators. Take the crash as an example, in most cases of crashes with app screenshots, there are such indicators in the app screenshots. It is the same with the performance cases, the reports with app screenshots submitted by crowdworkers are usually with the indicating icons. Generally speaking, we design the consistency detection strategies with the combination of external resource reference and the crowdsourced testing scenario consideration to ensure the strategy generalizability.}

\begin{equation}
	\label{equ:top3}
	S_{top-i}^* = max\{\delta_i * S_{dt}(top-i)\}, i = 1, 2, 3
\end{equation}

\begin{equation}
	\label{equ:score}
	res = 
	\begin{cases}
		1 & S_{top-3}^* \geq \lambda \\
		0 & otherwise
	\end{cases}
\end{equation}

As illustrated in \secref{sec:classifier}, we obtain 3 possible types from \classifier with top-3 confidence. However, due to the differences in confidence, we assign different weights for different results, $\delta_i$ for $top-i$ result. In our implementation of \toolname, we assign $\delta_1$, $\delta_2$, and $\delta_3$ to 1, 0.9, and 0.8, respectively, which shows the best performance according to practical experience and preliminary evaluation \vv{(details of preliminary evaluation shown in \figref{fig:config})}. As shown in \equref{equ:top3}, the $S_{top-3}^*$ is the maximum value among the three productions of $\delta_i$ and $S_{dt}(top-i)$, and if $S_{top-3}^*$ is greater than the preset threshold $\lambda$, \detector will confirm the consistency of the crowdsourced test report (\vv{\equref{equ:score}}). The $\lambda$ is set to 0.5 based on the experience from the industry.

\detector designs the \textit{General Strategy} and the \textit{Specific Strategies} to deal with different types of bugs. To fulfill these strategies, we build the locating feature dataset, including color words, position words, text words, and type words. Also, semantic identifying datasets are used to assist the \detector, including negative words and double-negative words. With the keyword dataset and the extracted information from app screenshots and textual descriptions with \decomposer, \detector is capable of detecting the mobile app crowdsourced test report consistency.

\section{Evaluation}
\label{sec:exp} 

In this section, we in detail elaborate the empirical survey and the evaluation of the effectiveness of \toolname. The empirical study indicates the necessity of this work to improve the crowdsourced testing quality, and the experiment results show the effectiveness of \toolname in detecting consistency in mobile app crowdsourced test reports.

\subsection{Experimental Setup}

\subsubsection{\textbf{Research Question}} 

We propose three research questions (RQ) to evaluate the effectiveness of the proposed approach.

\textbf{RQ1 (\classifier Effectiveness)}: How effective can \toolname classify the bugs in crowdsourced test reports?

\textbf{RQ2 (\toolname Effectiveness)}: How effective can \toolname detect the crowdsourced test report consistency?

\textbf{RQ3 (\toolname Usefulness)}: How useful can \toolname assist app developers in reviewing crowdsourced test reports in practical scenarios?

\subsubsection{\textbf{Data Collection and Preprocessing}}
\label{sec:data}

In this paper, we collect \dataNum mobile app crowdsourced test reports from the MoocTest crowdsourced testing platform\footnote{\url{http://www.mooctest.net}}, which is one of the representative and most popular crowdsourced testing platforms in China \vv{and has supported many academic studies in crowdsourced testing \cite{feng2016multi} \cite{yu2021prioritize} \cite{gao2019successes} \cite{feng2015test} \cite{yu2019crowdsourced}. According to the statistics provided in March 2021, the MoocTest platform has attracted over 20,000 users from 50 different countries, showing the great popularity and typicality of the platform.}
\vvvvv{Besides, in order to better illustrate the representativeness of MoocTest, we have an in-depth investigation into other platforms, like Test IO, Global App Testing, Baidu Crowdtesting, Testin, etc. We find that such platforms adopt almost identical crowdsourced testing mechanisms and involve similar crowdsourced work capabilities. For MoocTest and other platforms, there are two kinds of participants, the requesters, and the crowdworkers. The requesters publish their testing requirements as tasks on the platforms and attach the apps under test. Then crowdworkers choose to execute tasks with their own devices. If they find bugs, they will submit crowdsourced test reports, which consist of app screenshots and textual descriptions. In the textual descriptions, crowdworkers will state the steps triggering the bugs they find from the app launch to the bug occurrence and the descriptions of the bugs. Then the requesters can download all the reports within a fixed time period from crowdworkers. The capability of crowdworkers can affect the test reports. For MoocTest and other platforms, the crowdworkers are mixed with common end users and professional testers, and most of them are common end users. They may not precisely describe bugs in the apps under test. The crowdworkers may execute crowdsourced testing tasks on different platforms. Therefore, we believe that the almost identical crowdsourced testing mechanism and similar crowdsourced work capability involvement lead to the commonality of the inconsistency report problem.}

The test reports collected from the platform are of over 50 different mobile apps, and over 1,100 crowdworkers participate in the crowdsourced testing tasks. In order to label the consistency or inconsistency of the crowdsourced test reports, we recruit five participants to label the test reports. The five involved participants are graduate students or senior testing engineers with over three years of experience in mobile app testing and are familiar with crowdsourced testing. The participants spend three months in total manually labeling and cross-validating the data. 

Specifically, during the labeling work and the cross-validation, the participants are required to label the consistency or inconsistency of the test reports first, and the results should be the consensus of all five participants after discussion. \vv{The participants are instructed with the criterion of identifying a consistent or inconsistent report, which is to judge whether one of the bugs revealed in app screenshots can match one of the bugs described in textual descriptions. The Fleiss' kappa value of the five participants is around 0.95, which means ``almost perfect agreement''.}  Second, two of the participants label the bug type of each test report separately according to the pre-defined taxonomy. If their opinions are the same, the bug type would be confirmed. Otherwise, if the results are different, the other three participants will vote for the final results. Besides, a discussion among all five participants would be held for the bugs that are hard to determine the types. \vv{During the bug category labeling work, the Fleiss' kappa value is around 0.88, which means ``almost perfect agreement''.} Third, the participants will label the locating features in the textual descriptions. Each test report can be labeled with one or more locating features. Besides, the participants label the existence of negative words, double-negative words, and prompt words. \vv{To label the locating features, we accept the opinions of different participants to fully explore the keywords in textual descriptions to form a final set. The Fleiss' kappa value is around 0.9 in reaching the consensus of the final keyword sets.}

\subsection{RQ1: \classifier Effectiveness}

\begin{table}[!htbp]
\centering
\caption{\vv{Test Report Classification Result}}
\scalebox{1}{
\begin{tabular}{c|c|c|c|c|c}
\toprule

Data & Metric & Approach & top-1 & top-2 & top-3 \\ \midrule

\multirow{16}{*}{\rotatebox{90}{Test Set}} 

& \multirow{4}{*}{accuracy} 
&   SETU          & 30.74\% & 48.50\% & 48.50\% \\  
& & Bert (org)    & 10.43\% & 17.63\% & 25.12\% \\   
& & Bert (no-aug) & 56.60\% & 55.47\% & 62.68\% \\   
& & Bert (incon)  & 72.87\% & 90.31\% & 96.16\% \\ 
& & \bgg{ReCoDe-Bert}  & \bgg{80.78\%} & \bgg{94.22\%} & \bgg{96.91\%} \\ 
\cmidrule{2-6} 

& \multirow{4}{*}{precision} 
&   SETU          & 35.17\% & 44.88\% & 44.91\% \\  
& & Bert (org)    & 2.54\%  & 12.32\% & 20.06\% \\  
& & Bert (no-aug) & 31.66\% & 54.66\% & 71.62\% \\   
& & Bert (incon)  & 74.97\% & 88.82\% & 92.16\% \\  
& & \bgg{ReCoDe-Bert}  & \bgg{81.22\%} & \bgg{95.23\%} & \bgg{97.10\%} \\ 
\cmidrule{2-6} 
 
& \multirow{4}{*}{recall}
&   SETU          & 44.87\% & 64.91\% & 64.92\% \\  
& & Bert (org)    & 10.36\% & 21.71\% & 31.41\% \\  
& & Bert (no-aug) & 45.36\% & 58.71\% & 61.41\% \\   
& & Bert (incon)  & 73.16\% & 85.14\% & 91.93\% \\  
& & \bgg{ReCoDe-Bert}  & \bgg{82.74\%} & \bgg{94.70\%} & \bgg{97.23\%} \\ 
\cmidrule{2-6} 

& \multirow{4}{*}{F1 score}
&   SETU          & 39.43\% & 53.07\% & 53.09\% \\  
& & Bert (org)    & 4.08\%  & 15.72\% & 24.48\% \\  
& & Bert (no-aug) & 37.29\% & 56.61\% & 66.12\% \\   
& & Bert (incon)  & 74.05\% & 86.94\% & 92.04\% \\  
& & \bgg{ReCoDe-Bert}  & \bgg{81.97\%} & \bgg{94.97\%} & \bgg{97.17\%} \\ 
\midrule

\multirow{16}{*}{\rotatebox{90}{All Report Set}} 

& \multirow{4}{*}{accuracy}  
&   SETU          & 27.66\% & 45.52\% & 56.57\% \\  
& & Bert (org)    & 4.70\%  & 7.79\%  & 7.86\%  \\  
& & Bert (no-aug) & 46.36\% & 54.67\% & 61.78\% \\    
& & Bert (incon)  & 71.82\% & 89.01\% & 94.78\% \\ 
& & \bgg{ReCoDe-Bert}  & \bgg{90.98\%} & \bgg{96.95\%} & \bgg{98.76\%} \\ 
\cmidrule{2-6} 

& \multirow{4}{*}{precision} 
&   SETU          & 35.80\% & 43.27\% & 51.59\% \\  
& & Bert (org)    & 3.62\%  & 6.83\%  & 6.80\%  \\  
& & Bert (no-aug) & 38.66\% & 61.66\% & 75.62\% \\    
& & Bert (incon)  & 65.49\% & 81.23\% & 87.31\% \\ 
& & \bgg{ReCoDe-Bert}  & \bgg{92.03\%} & \bgg{95.27\%} & \bgg{95.92\%} \\ 
\cmidrule{2-6} 

& \multirow{4}{*}{recall}    
&   SETU          & 39.73\% & 60.27\% & 70.22\% \\  
& & Bert (org)    & 10.11\% & 20.53\% & 20.68\% \\  
& & Bert (no-aug) & 44.11\% & 55.53\% & 57.68\% \\    
& & Bert (incon)  & 71.12\% & 83.01\% & 89.78\% \\ 
&& \bgg{ReCoDe-Bert}  & \bgg{93.66\%} & \bgg{96.50\%} & \bgg{97.67\%} \\ 
\cmidrule{2-6} 

& \multirow{4}{*}{F1 score}
&  SETU          & 37.66\% & 50.38\% & 59.48\% \\  
& & Bert (org)    & 5.33\%  & 10.25\% & 10.23\% \\  
& & Bert (no-aug) & 41.21\% & 58.43\% & 65.44\% \\   
& & Bert (incon)  & 68.19\% & 82.11\% & 88.53\% \\  
& & \bgg{ReCoDe-Bert}  & \bgg{92.83\%} & \bgg{95.88\%} & \bgg{96.79\%} \\ 
\bottomrule

\end{tabular}}
\label{tbl:classifier}
\end{table}

\begin{table*}[!htbp]
\centering
\caption{\vv{Breakdown Performance of the \classifier}}
\scalebox{1}{
\begin{tabular}{c|c|c|c|c|c|c|c|c|c|c}
\toprule

\diagbox{Fact}{Prediction}
& \textit{\tabincell{c}{Garbled \\ Error}}     
& \crash     
& \textit{\tabincell{c}{Display \\ Problem}}     
& \textit{\tabincell{c}{Functional \\ Defect}} 
& \textit{\tabincell{c}{Layout \\ Problem}} 
& \textit{\tabincell{c}{Network \\ Error}} 
& \textit{\tabincell{c}{Performance \\ Problem}}  
& \textit{\tabincell{c}{Null \\ Screen}}   
& \textit{\tabincell{c}{Transition \\ Problem}} 
& \textit{\tabincell{c}{Error \\ Prompt}}  \\ \midrule

\garerr & \bgg{86.6\%} & 0.0\%  & 0.7\%  & 10.7\% & 0.7\%  & 0.7\%  & 0.0\%  & 0.0\%  & 0.0\%  & 0.7\%  \\
\crash  & 0.0\%  & \bgg{98.9\%} & 0.2\%  & 0.7\%  & 0.0\%  & 0.0\%  & 0.2\%  & 0.0\%  & 0.0\%  & 0.0\%  \\
\dispro & 0.1\%  & 0.0\%  & \bgg{79.7\%} & 16.5\% & 2.2\%  & 0.2\%  & 0.5\%  & 0.0\%  & 0.5\%  & 0.2\%  \\
\func   & 0.0\%  & 0.1\%  & 1.1\%  & \bgg{98.0\%} & 0.1\%  & 0.1\%  & 0.2\%  & 0.0\%  & 0.1\%  & 0.3\%  \\
\laypro & 0.0\%  & 0.0\%  & 1.7\%  & 0.7\%  & \bgg{97.5\%} & 0.0\%  & 0.0\%  & 0.0\%  & 0.1\%  & 0.0\%  \\
\neterr & 0.1\%  & 0.4\%  & 0.4\%  & 2.8\%  & 0.1\%  & \bgg{94.8\%} & 0.3\%  & 0.0\%  & 0.4\%  & 0.6\%  \\
\perfp  & 0.0\%  & 0.1\%  & 0.3\%  & 0.5\%  & 0.0\%  & 0.5\%  & \bgg{98.4\%} & 0.0\%  & 0.1\%  & 0.1\%  \\
\nullsr & 0.0\%  & 0.6\%  & 3.8\%  & 2.5\%  & 0.2\%  & 0.2\%  & 0.6\%  & \bgg{91.5\%} & 0.2\%  & 0.6\%  \\
\tranpr & 0.0\%  & 0.2\%  & 1.2\%  & 2.5\%  & 0.2\%  & 0.0\%  & 0.4\%  & 0.1\%  & \bgg{95.3\%} & 0.2\%  \\
\errorp & 0.3\%  & 0.2\%  & 1.3\%  & 16.1\% & 0.1\%  & 1.3\%  & 0.6\%  & 0.0\%  & 0.5\%  & \bgg{79.5\%} \\ \bottomrule
\end{tabular}}
\label{tbl:breakdown}
\end{table*}

In this research question, we research the \classifier capability. \classifier is one important basis for the whole \toolname, and the classifying results determine the strategies applied on the test reports for consistency detection. Moreover, in this RQ, \augmentor is also evaluated as a significant basis for \classifier.

The \classifier solves a multiclass classification problem. As commonly defined, a true positive prediction (TP) is correctly predicting a positive sample, and a false positive prediction (FP) is incorrectly predicting a positive sample. A true negative prediction (TN) is correctly predicting a negative sample, and a false negative prediction (FN) is incorrectly predicting a negative sample. We use the $accuracy$, $precision$, $recall$, and $F1~score$ values to evaluate the effectiveness of \classifier. The values are calculated as: $accuracy = \frac{TP + TN}{TP + NP + TN + FN}$, $precision = \frac{1}{n} \sum_{i=1}^{n} \frac{TP_i}{TP_i + FP_i}$, $recall = \frac{1}{n} \sum_{i=1}^{n} \frac{TP_i}{TP_i + FN_i}$, and $F1~score = \frac{1}{n} \sum_{i=1}^{n} \frac{2 \times precision_i \times recall_i}{precision_i + recall_i}$.

We choose one of the state-of-the-art crowdsourced test processing approaches, SETU \cite{wang2019images}, as a baseline of the \classifier (denoted as \bert). \vv{The general idea of SETU is to extract features from both app screenshots and textual descriptions (we use the textual description feature part as the baseline). SETU is a state-of-the-art tool for crowdsourced test report processing, including report classification. The most important part of report classification is to effectively extract features from the crowdsourced test reports. SETU is one representative study that can effectively extract features for report classification. Further, another important factor for report classification is the taxonomy. For other advanced tools for general bug report classification, the taxonomy may be completely different, while for SETU, which also focuses on the crowdsourced testing scenario, the taxonomy is consistent with \toolname. Moreover, SETU utilizes TF-IDF and word2vec features of textual descriptions, which are quite classic and widely used features in the NLP field.} 

\vv{In order to show the effectiveness of \bert from different perspectives, other different configurations of the \bert are discussed as baselines, including the pre-trained BERT model in \cite{devlin2018bert} (labeled as BERT (org) in \tabref{tbl:classifier}), the BERT model trained with a non-augmented dataset with imbalanced data distribution (labeled as BERT (no-aug) in \tabref{tbl:classifier}), and the BERT model trained with a mixture of inconsistent crowdsourced test reports (labeled as BERT (incon) in \tabref{tbl:classifier}).}

\vv{The training (fine-tuning) time of the BERT model is about four hours with a Tesla V100 SXM2 32GB GPU. Our machine is with a 10-core Intel Xeon Gold 6248 CPU @ 2.50GHz and 108 GB of memory. For the report classification, the approximate time is 1.03 seconds for predicting the category for each test report.}

The results of RQ1 are shown in \tabref{tbl:classifier}. The \bert model is trained on the augmented consistent test report dataset (\secref{sec:augmentor}). The dataset contains 3,740 crowdsourced test reports and is divided into the training set, validation set, and test set at the ratio of 6:2:2, which is the common practice of text classification tasks. We set the batch size as 32, and train the model for 30 epochs. \vv{All the hyper-parameters are kept the same in \cite{devlin2018bert}.}

\tabref{tbl:classifier} (metrics labeled with ``\textit{Test Set}'') shows the final results on the test set after the model training. For the top-3 results of \bert, the $accuracy$ reaches 96.91\%, the $precision$ reaches 97.10\%, the $recall$ reaches 97.23\%, and the $F1~score$ reaches 97.17\%. 
Compared with the baseline approach, SETU, \toolname outperforms by 99.81\%, 116.21\%, 49.77\%, and 83.01\% for four metrics, respectively. It shows using the BERT model trained with crowdsourced test report data, the effectiveness can be greatly improved.
\vv{The effectiveness of the BERT (org) model is quite poor, and the reason is that the BERT (org) model is trained with data in common scenarios. Such a BERT (org) model cannot effectively extract the features of the texts in the crowdsourced testing scenario.
Compared with the BERT (no-aug) model, \bert outperforms by 54.61\%, 35.57\%, 58.34\%, and 46.95\% for four metrics, respectively. This means that though the model trained with dataset without \augmentor can capture the features of the textual descriptions describing bugs of different categories, it is still influenced by the imbalanced data distribution problem, which weakens the BERT (no-aug) model's capability to classify the textual descriptions into correct categories.
Compared with the BERT (incon) model, \bert outperforms by 0.78\%, 5.36\%, 5.77\%, and 5.56\% for four metrics, respectively. Such a result shows that whether the training dataset contains textual descriptions from inconsistent reports or not, the effectiveness is not affected much. No matter whether the reports are consistent or not, the textual descriptions can provide information for \bert (or BERT (incon) model) to classify crowdsourced test reports.}
Besides, the values of the top-1 results are all over 80\%, and the values of the top-2 results reach approximately 95\%. \bert's top-1 and top-2 results perform much better than the baseline and other configurations of the BERT models. The results show that \bert performs quite well in classifying the crowdsourced test reports into corresponding bug types.

\begin{table*}[!htbp]
	\centering
	\caption{\vv{\toolname Effectiveness}}
	\scalebox{1}{
	\begin{tabular}{c|c|c|c|c|c|c|c|c|c}
	\toprule
	\multirow{2}{*}{Metric} & 
	\multicolumn{3}{c|}{\toolname} & 
	\multicolumn{2}{c|}{\multirow{2}{*}{ViLBERT}} &
	\multicolumn{2}{c|}{\multirow{2}{*}{\toolname (one stage)}} &
	\multicolumn{2}{c}{\multirow{2}{*}{SETU + \detector}} \\ \cline{2-4}
	& top-1   & top-2   & top-3 & 
	\multicolumn{2}{c|}{} & \multicolumn{2}{c|}{}& \multicolumn{2}{c}{} \\ \midrule
	
	accuracy  & 86.10\% & 89.54\% & \textbf{91.35\%} & 42.10\% & $\uparrow$ 116.98\%  & 59.29\% & $\uparrow$ 54.07\% & 51.50\% & $\uparrow$ 77.37\% \\
	precision & 57.73\% & 65.72\% & \textbf{70.39\%} & 42.10\% & $\uparrow$ 67.20\% & 51.89\% & $\uparrow$ 35.65\%  & 35.24\% & $\uparrow$ 99.75\%  \\
	recall & 86.07\% & 87.99\% & \textbf{89.94\%} & \st{100.00\%} & \st{$\uparrow$ -10.06\%} & 81.83\% & $\uparrow$ 9.91\% & 84.87\% & $\uparrow$ 5.97\%  \\
    F1 score  & 69.11\% & 75.24\% & \textbf{78.97\%} & 59.25\% & $\uparrow$ 33.29\%  & 63.51\% & $\uparrow$ 24.35\% & 49.80\% & $\uparrow$ 58.58\% \\ \bottomrule
	\end{tabular}}
	\label{tbl:detector}
\end{table*}

Further, we have experimented with extended data. \tabref{tbl:classifier} (metrics labeled with ``\textit{All Report Set}'') shows the generalization capability of the \classifier. The data under experiment include all test reports of consistent and inconsistent ones, with the meaningless reports being eliminated (meaningless reports refer to the reports that are not describing bugs, and only submit textual descriptions of meaningless texts, \eg ``11111111...'' \cite{yu2019crowdsourced}). \vv{The extended data are directly fed into the \classifier, which is the same one that is used to evaluate on the ``Test Set'' data, and the output is the corresponding predicted category to evaluate the effectiveness of the classifier. The extended data (``All Report Set'') are not used in the training process.} One thing to notice is that even though textual descriptions in inconsistent reports may not match the app screenshots, they still express the bugs, so we evaluate the \bert on such reports. For the extended data, the top-3 $accuracy$ reaches 98.76\%, the $precision$ reaches 95.92\%, the $recall$ reaches 97.67\%, and the $F1~score$ reaches 96.79\%. \vv{\classifier outperforms the baseline approach and different configurations of the BERT model (including the BERT (org) model, the BERT (no-aug) model, and the BERT (incon) model). The improvement on most results on the ``All Report Set'' is even more obvious than the results on the ``Test Set'', which shows that \bert not only has better effectiveness in classifying the textual descriptions to different bug categories but also has much better generalizability.} 

A seemingly counter-intuitive phenomenon is that \classifier (labeled as \bert in \tabref{tbl:classifier}) performs even better on extended data than on the test set. We investigate and find that the data in the test set distribute equally in 10 bug types, while the data distribution of the extended dataset is different because they are collected from industry, and some types of bugs may appear more frequently, \eg \crash, \nullsr (see \secref{sec:survey}). Such problems may have more apparent features for the \classifier. Also, \toolname has a strong capability to extract features. The experiment results indicate that \classifier of \toolname has good generalization capability, and can accurately classify the test reports to the corresponding bug type according to the textual descriptions.

\vv{In order to better illustrate the performance of \classifier in detail, we also present the breakdown performance on each of the ten categories in \tabref{tbl:breakdown}. The results further show the good performance of \classifier in classifying the 10 categories of bugs presented in crowdsourced testing.}

\subsection{RQ2: \toolname Effectiveness}

The \detector can be seen as a binary classification problem, the consistent reports are labeled 1, and the inconsistent ones are labeled 0, so we use the $accuracy$, $precision$, $recall$, and $F1~score$ values to evaluate the \toolname. The calculation is as: $precision = \frac{TP}{TP + FP}$, $recall = \frac{TP}{TP + FN}$, and $F1~score = \frac{2 \times precision \times recall}{precision + recall}$. Also, \decomposer is evaluated for decomposing the app screenshots and textual descriptions for the \detector.

\begin{figure}[!htbp]
    \centering
    \includegraphics[width=\linewidth]{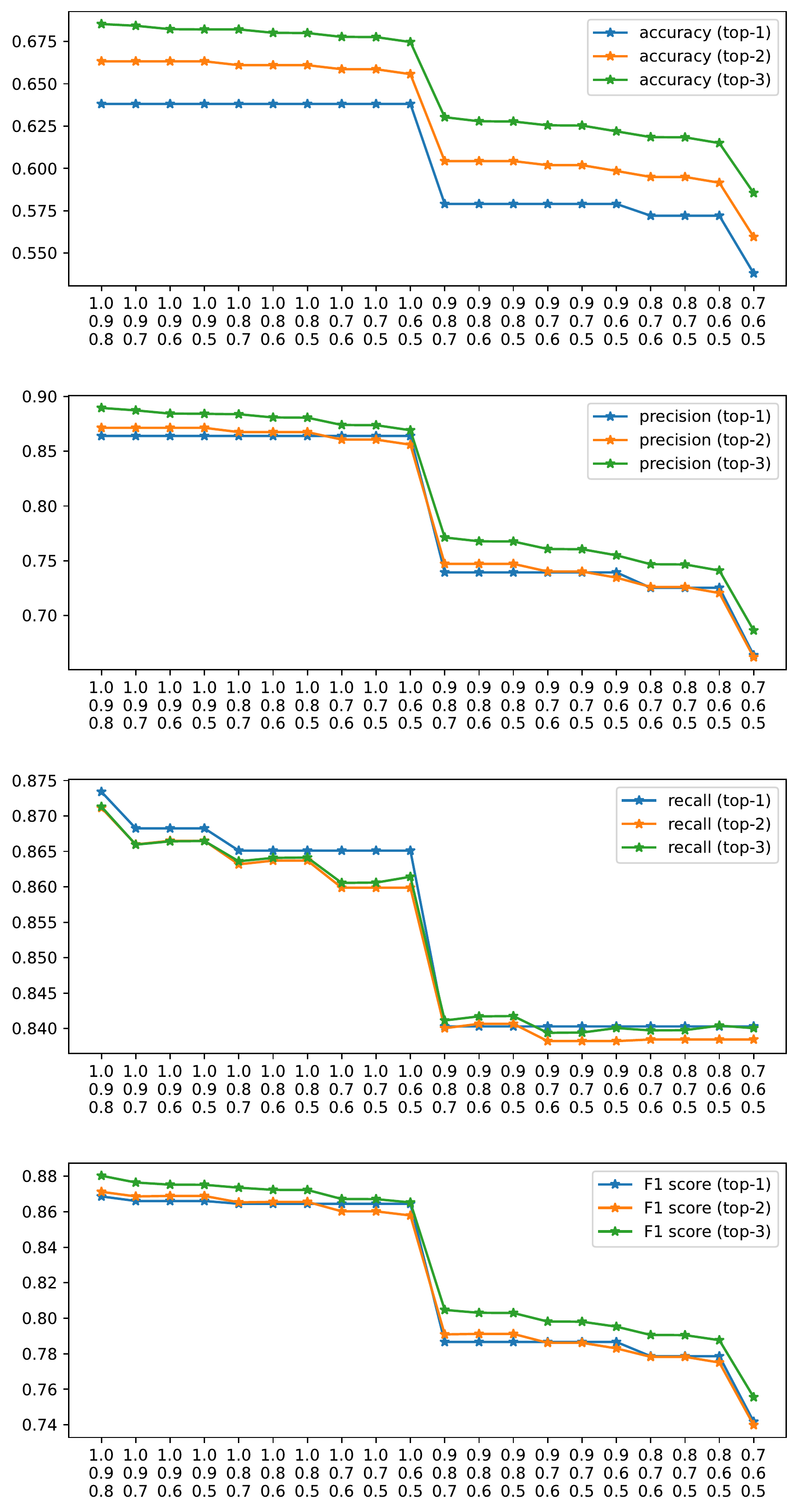}
    \caption{\vv{Effectiveness Comparison of different $\delta_i$ Configuration}}
    \label{fig:config}
\end{figure}

First, the $\delta_i$ is a set of critical parameters in the \detector construction. In order to confirm the optimized configuration of $\delta_i$, we conduct a preliminary evaluation on a sub-dataset (\figref{fig:config}).

\vvvvv{The dataset in the RQ2 evaluation contains all the crowdsourced test reports that are not used in the training of the \classifier of \toolname to avoid the data snooping problem. The dataset contains 19,035 crowdsourced test reports, excluding the augmented reports and only considering real-world reports. Besides, during RQ2, we do not provide perfect report type labels for each test report but use the \classifier to identify the types. This practice is to evaluate the \toolname as a whole.}

To evaluate the effectiveness of \toolname, we set three baselines. The first is one of the state-of-the-art one-stage approaches, ViLBERT\cite{lu2019vilbert}, which is originally used for general scene image and text consistency detection. ViLBERT is a representative one-stage approach that is based on deep learning models \cite{li2020unicoder} \cite{su2019vl}. During the retraining with the test report dataset, we keep the original settings and hyperparameters and construct the training set as the original dataset. The model is trained for 50 epochs. The second (denoted as \toolname (one stage)) is the single \detector of \toolname. In other words, we apply the \textit{General Strategy} to all test reports to identify the report consistency. The third baseline (denoted as SETU+\detector) is to replace the \toolname \classifier with SETU \cite{wang2019images}, one of the state-of-the-art approaches for test report classification.

\tabref{tbl:detector} show the experiment results. 
For the $accuracy$ value, \toolname (top-3) reaches 91.35\%, and compared with baselines, \toolname outperforms by \textbf{116.98\%}, \textbf{54.07\%} and \textbf{77.37\%} respectively. Moreover, \toolname (one stage) also performs better than the ViLBERT. As for the $precision$ value, \toolname (top-3) reaches 70.39\%, outperforming baselines by \textbf{67.20\%}, \textbf{35.65\%} and \textbf{99.75\%} respectively. \vv{The $recall$ of \toolname (top-3) reaches 89.94\%, outperforming the baselines (except ViLBERT) by \textbf{9.91\%} and \textbf{5.97\%}, respectively. $F1~score$ of \toolname (top-3) reaches 78.97\%, outperforming the baselines by \textbf{33.29\%}, \textbf{24.35\%} and \textbf{58.58\%}, respectively.} Experiment results of the $recall$ value of ViLBERT are abnormal and reach 100\%. We look insight the detailed data and find that \textbf{ViLBERT predicts all crowdsourced test reports as consistent}. Therefore, we hold that such models have no practical application value. Also, we infer that the reasons are that the data volume is far from enough for the model to extract features, and the widgets with bugs are non-salient, which leads to obstacles for the models to extract features. Generally speaking, \toolname has the best performance, and the \toolname (one stage) also performs well, which indicates the effectiveness of the designed detecting strategies. SETU+\detector results show that \classifier performs much better than SETU.

\subsection{\vv{RQ3: User Study}}

\vv{

In order to better illustrate the practical use of the proposed \toolname, we design and implement a user study. The user study is divided into two parts with 10 participants. In the first part, the 10 participants are required to experience the bug inspection process with and without \toolname. This part aims to quantificationally evaluate how \toolname can improve the bug reviewing process. In the second part, the 10 participants are required to experience \toolname and answer a questionnaire. This part aims to collect the assessment from the actual users of \toolname.

\subsubsection{User Study Setting}

We totally recruit 10 people to participate in the user study. Such participants are all senior software engineering graduate students with more than three years of Android testing and crowdsourced testing experience in the real industry environments. According to Salmon \etal \cite{salman2015students}, experienced graduate students are sufficient developer proxies under carefully controlled experiments, which means that the participants can actually represent real industry developers. The 10 students are all Chinese. Six of them are male and four are female. Three are fourth-year undergraduate students and seven are master students. 

\begin{figure}[!htbp]
    \centering
    \includegraphics[width=0.7\linewidth]{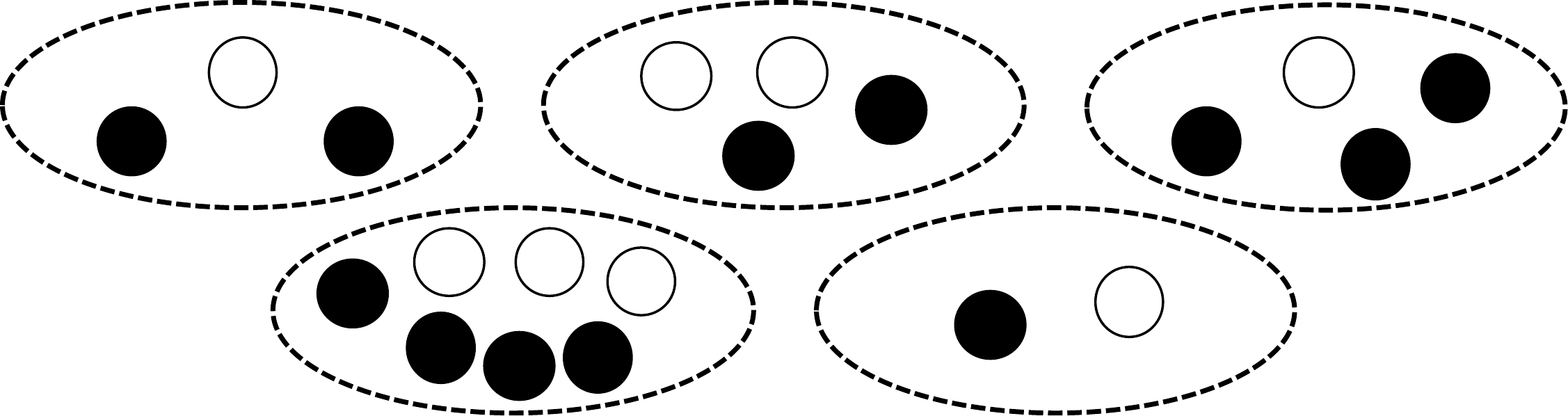}
    \caption{\vv{User Study of Using \toolname}}
    \label{fig:ussetting}
\end{figure}

For the app used in the user study, we use a popular app on GitHub, and the participants are all familiar with the developing framework and technology of the app. On the day before the task, the participants are given the app installing package to get familiar with. The app is with 20 crowdsourced test reports, among which eight are consistent. The reports contain 5 bugs. In \figref{fig:ussetting}, each group means a bug. The white circle represents consistent reports, and the black represents inconsistent ones.

\subsubsection{Bug Inspection Efficiency Comparison}

\begin{figure}[!htbp]
    \centering
    \includegraphics[width=\linewidth]{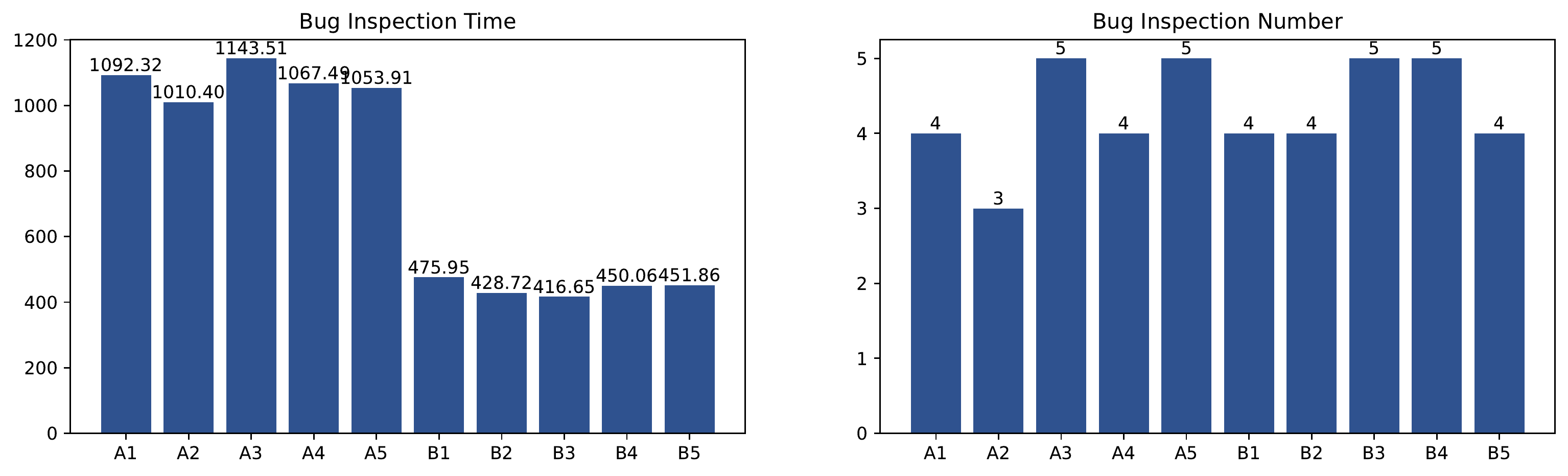}
    \caption{\vv{User Study of Using \toolname}}
    \label{fig:userstudy}
\end{figure}

In the bug inspection efficiency comparison, we divide the 10 participants into two groups. Participants in the first group are required to review all the reports without any assistance from \toolname, and participants in the second group are with the results of \toolname, and they can skip the reports labeled as inconsistency. In order to use \toolname, the participants only need to upload the directory that stores the crowdsourced test reports, and the \toolname can automatically generate the result CSV file for all the reports. The participants are required to review the crowdsourced test reports and reproduce the bugs. We perform the Friedman test to evaluate the participants' capability difference, and the significance value $p$ is calculated as 0.031, which is less than 0.05, showing that the capability is with no significant difference. The time for bug inspection and reproduction is recorded to show the efficiency. According to \figref{fig:userstudy}, we can find that the average time to review all required reports of the first group is 1073.53 seconds, and the average of the second group is 444.65 seconds. The results indicate that the efficiency is improved by 1.4 times ($\frac{1/444.65-1/1073.53}{1/1073.53} = 1.414$). We also calculate the bugs found by each participant. The participants in the first group on average find 4.2 (out of 5) bugs, and the participants in the second group on average find 4.4 (out of 5) bugs, indicating that using \toolname to eliminate inconsistent test reports can keep the bug inspection effectiveness.
Generally speaking, the results above show that \toolname can actually improve the bug inspection efficiency with the bug inspection effectiveness being kept.

\subsubsection{Questionnaire}

\begin{table*}[!htbp]
\centering
\caption{\vv{Questionnaire for User Study}}
\scalebox{1}{
\begin{tabular}{l|c|c|c|c|c|c|c|c|c|c|c|c}
\toprule
\textbf{Question} & \textbf{Goal} & \textbf{U1} & \textbf{U2} & \textbf{U3} & \textbf{U4} & \textbf{U5} & \textbf{U6} & \textbf{U7} & \textbf{U8} & \textbf{U9} & \textbf{U10} & \textbf{\textit{Avg.}} \\ \midrule
Whether ReCoDe is easy to use for you? & Usability     & 4 & 5 & 5 & 4 & 5 & 4 & 5 & 5 & 5 & 4 & 4.6 \\ \midrule
\tabincell{l}{Do you think extra tutorial or training are required \\ before you use ReCoDe?} & Accessibility & 5 & 5 & 4 & 5 & 5 & 5 & 4 & 5 & 4 & 5 & 4.7 \\ \midrule
\tabincell{l}{Can ReCoDe provide clear results for you about \\ the report consistency?} & Usefulness    & 5 & 4 & 5 & 5 & 5 & 4 & 5 & 5 & 5 & 4 & 4.7 \\ \midrule
\tabincell{l}{Are the results provided by ReCoDe correspond to \\ your opinions to the reports?} & Correctness   & 5 & 4 & 5 & 4 & 5 & 5 & 5 & 4 & 5 & 4 & 4.6 \\ \midrule
Can ReCoDe reduce your time during bug reviewing? & Efficiency    & 5 & 4 & 5 & 4 & 5 & 5 & 5 & 4 & 5 & 5 & 4.7 \\ \midrule
\tabincell{l}{Can ReCoDe reduce your work burden during the \\ bug reviewing process?} & Helpfulness   & 4 & 5 & 5 & 5 & 5 & 4 & 5 & 5 & 5 & 5 & 4.8 \\ \bottomrule
\end{tabular}}
\label{tbl:qa}
\end{table*}

We design a questionnaire for all the participants with the goal to show how actually do the participants feel about \toolname. Before answering the questionnaire, all the participants can view all the report labeling results of \toolname. There are six questions in the questionnaire with different goals to evaluate the \toolname from distinct perspectives, including Usability, Efficiency, Usefulness, Correctness, Helpfulness, and Accessibility, \vvvvv{following the System Usability Scale}. \vvvvv{The questions are Likert-based questions and all} the participants are required to give scores from 1 to 5, where 1 indicates a bad tool and 5 indicates a good tool. \vvvvv{Specifically, in order to eliminate the bias of the questions, we instruct the participants about the scores for each question. For question 1, score 1 indicates the tool is very difficult to use and score 5 indicates the tool is very easy to use. For question 2, score 1 indicates the participants need huge efforts of training before the tool usage and score 5 indicates the participants need few efforts of training before the tool usage. For question 3, score 1 indicates the tool can only provide very vague results and score 5 indicates the tool can provide very clear results. For question 4, score 1 indicates the results of the tool correspond to the participants' opinions very badly and score 5 indicates the results of the tool correspond to the participants' opinions very well. For question 5, score 1 indicates the tool can hardly reduce the reviewing time and score 5 indicates the tool can greatly reduce the reviewing time. For question 6, score 1 indicates the tool can hardly reduce the reviewing work burden and score 5 indicates the tool can greatly reduce the reviewing work burden.} From the \tabref{tbl:qa}, the results show that most participants are satisfied with the tool. They think \toolname is easy to use and user-friendly even for new comers of this tool, because they only need to identify the path where the crowdsourced test report files are stored, \toolname will automatically generate the consistency labels for each report for their reference.  Besides, \toolname can provide correct and clear results for app developers to review the crowdsourced test reports, which is a useful guidance to them. Further, all the participants agree that \toolname can improve bug reviewing efficiency and relieve their work burden due to the elimination of inconsistent crowdworkers, and the bugs are not lost.

}

\subsection{Threats to Validity}

In this section, we introduce the possible threats and our actions to eliminate such threats.

\textbf{Internal Threats.}
The internal threats to validity are the \toolname implementation and the model adaptation. To migrate such threats, we invite a third-party group to review our implementation and the retraining of the original deep learning models. Therefore, the negative effects of code implementation are minimized.

\textbf{External Threats.}
One main external threat to validity is the representativeness of the apps under crowdsourced testing on the platform. To migrate the external threat, we use apps of different categories, including \textit{system}, \textit{internet}, \textit{tool}, \textit{music}, \textit{phone \& SMS}, \textit{finance}, \textit{development}, \textit{sports}, \textit{shopping}, and \textit{map}. The wide range of app categories can eliminate the external threat, and show the excellent generalization capability of the \toolname.

\textbf{Construct Threats.}
The ground truth is manually labeled, which may bring construct threats. However, we have tried to eliminate errors. First, the labeled results are double-checked and are consensus after discussion. Second, labeling participants are senior software engineers or experienced graduate students. As presented in \cite{salman2015students}, senior students are eligible for developer proxies in the controlled experiments.

\subsection{Discussion}
\label{sec:discuss}

\subsubsection{Significance of Report Consistency}

In this paper, we propose \toolname to detect the crowdsourced test report consistency. The intuition of this approach is on the basis of the large scale of inconsistent reports caused by the openness of crowdsourced testing. As an approach to improve the efficiency of crowdsourced testing, we hold that the elimination would help app developers reduce the number of crowdsourced test reports they need to review, and would promote the efficiency of app testing. 

Existing studies focus on report classification \cite{wang2016local} \cite{wang2019images}, clustering or prioritization \cite{yu2021prioritize}, \etc However, such work mainly helps better organize all the submitted report, instead of filtering out the reports that have no effects or even negative effects. Therefore, we believe it is significant to propose an approach to help reduce the amount of reports to be reviewed, and meantime to make the important bugs can be kept for review. However, we do not claim that all the reports left would have a positive effect on the bug inspection. Consistency of crowdsourced test reports is the lowest line for the reports to have positive effects on bug inspection of app developers. We do not mean that the consistent test reports of \toolname can necessarily assist in bug finding. The purpose is to eliminate the inconsistent reports, so as to get rid of the negative effects introduced by such reports, especially the waste of time.

\vv{Generally speaking, inconsistent test reports are one kind of low-quality test reports. Even though some reports are consistent, they may be low-quality. For example, if the crowdworker is with low testing expertise, the submitted test report cannot always precisely describe what has happened, even if the bug on the app screenshots is actually reported in the textual descriptions. For the meaningless test reports mentioned in this paper, the bugs reported in textual descriptions are zero, so no one can be matched to any bugs in app screenshots. Such reports are considered inconsistent. For the reports with GUI design suggestions, there are also no bugs in the textual descriptions, so such reports are also considered inconsistent. From this perspective, the reported number 18.07\% is the actual consistent test report rate, which can actually have a negative effect on the app developers' bug reviewing process.}
\vvvvv{As a subset of low-quality reports, the concept of inconsistency is a very objective criterion that can be strictly followed in this paper. However, the low-quality concept may be more subjective. For example, in one report, the crowdworker describes the bug behavior but does provide the reproduction steps, making it hard to reproduce the bug. This report is a low-quality report because it cannot guide the app developers to reproduce and locate the bug, but the described bugs can match the screenshot, so it is definitely consistent. It is subjective for app developers to judge the quality of the crowdsourced test reports, so it is hard to define criteria for automated approaches to follow, but it is definite to judge the consistency based on our definition in \toolname.}

\subsubsection{One-Stage Approach Comparison}

As shown in the experiment results, we can find that our approach can perform better than the state-of-the-art one-stage approach. We also provide an analysis of the results. However, with the development of deep learning models or other learning-based approaches, we also think the models would have the capability to focus on the subtle objects (widgets) on mobile app screenshots, and therefore be more robust for detecting test report consistency.

\vv{Specifically, inconsistency reports are only a subset of low-quality reports, and low-quality reports can hardly be identified with current NLP techniques. For low-quality reports, NLP techniques cannot recognize them without any references, like keyword matching, or learning-based feature extraction. 
A completely constructed rule system is always required \cite{chaparro2019assessing} for some current approaches. Such approaches consider more about the styles and formats of crowdsourced test reports, and they neglect the consistency in contents, which are more important. In this paper, we have a deep analysis on the app screenshots and textual descriptions, which can identify the inconsistency in the contents of crowdsourced test reports instead of styles or formats.
Therefore, \toolname is an effective and state-of-the-art approach to detect inconsistency test reports, as one kind of low-quality test report.}

\section{Related Work}
\label{sec:rw}

\subsection{Crowdsourced Testing Quality Control}

Crowdsourced testing has become more and more popular. Different from traditional testing paradigms, crowdsourced testing distributes testing tasks to a large group of crowdworkers, who are more like common end users instead of testing professionals. Crowdsourced testing utilizes the dispersion of crowdworkers' locations, devices, operating systems, and testing ideas.  However, every coin has two sides, and the openness of crowdsourced testing cause severe quality control problem. 

Many pieces of research have been done to solve such a problem. Some starts from filtering the crowdworkers \cite{xie2017cocoon} \cite{cui2017multi} \cite{cui2017should} \cite{wang2020context}. the papers propose different algorithms for selecting crowdworkers based on crowdworker features and task features, in order to control crowdsourced testing quality from the perspective of participants. However, the participant is only one of the important factors that affect the crowdsourced testing quality. Even professional or highly matched crowdworkers may fail to submit effective reports due to economic reasons (platforms may reward workers according to the quantity of tasks they complete).

More work would focus on test report processing, \ie report classification, duplication detection, and report prioritization. Such work reorganizes the reports and has a better way to present reports to app developers.

Jiang \etal proposed TERFUR \cite{jiang2009adaptive}, using NLP algorithms to analyze and cluster test cases. 
Sun \etal \cite{sun2010discriminative} build a novel information retrieval model for detecting duplicate bug reports. 
Sureka \etal \cite{sureka2010detecting} introduced a model using the character n-gram for duplicate detection. 
Nguyen \etal introduced DBTM \cite{nguyen2012duplicate}, combining IR-based and topic-based features, to detect bug report duplication. 
Many other studies focus on the report deduplication based on textual information \cite{banerjee2012automated} \cite{prifti2011detecting} \cite{sun2011towards} \cite{zhou2012learning} \cite{huang2020quest}.
DRONE, proposed by Tian \etal \cite{tian2013drone}, is a machine learning-based approach to predict the test report priority by extracting and comparing different report features. 
Banerjee \etal \cite{banerjee2013fusion} proposed a multi-label classifier to find the ``primary'' report of a cluster of reports.
Alipour \etal \cite{alipour2013contextual} had a more comprehensive analysis of the test report context and improved the duplication detection accuracy. 
Wang \etal \cite{wang2016towards} consider the features of crowdworkers as a feature of test reports, and then complete the cluster task. 
They further \etal propose the LOAF \cite{wang2016local}, which is the first to separate operation steps and result descriptions for feature extraction. 
Hindle \cite{hindle2016contextual} makes improvements by combining contextual quality attributes, architecture terms, and system-development topics to improve deduplicate detection.
Feng \etal proposed a series of approaches, DivRisk \cite{feng2015test} and BDDiv \cite{feng2016multi}, to prioritize the test reports, and they first utilize the test reports screenshots. 
Yu proposed CroReG \cite{yu2019crowdsourced}, which can analyze the app screenshots and generate the corresponding textual descriptions that are explaining the bugs.
Wang \etal \cite{wang2019images} work further and explore a more sound approach for test report prioritization by paying more attention to app screenshots.
Yu \etal proposed DeepPrior \cite{yu2021prioritize}, a tool that mines image semantics and text semantics to assist in prioritizing crowdsourced test reports.

The aforementioned technologies mitigate the quality control problem in crowdsourced testing and improve the test report reviewing efficiency. However, app developers still have to review all test reports, including consistent and inconsistent ones. Consistent reports are still mixed with inconsistent ones, which will lead to confusion for app developers when they review and reproduce the bugs. This will waste app developers' time and effort.

\vv{Report quality control is important to the bug reproduction with test reports. Fazzini \etal \cite{fazzini2018automatically} propose Yakusu to generate executable test cases from bug reports with the help of a combination of program analysis and natural language processing techniques. Similarly, Zhao \etal \cite{zhao2022recdroid+} propose ReCDroid+ to automatically reproduce crashes from bug reports for Android apps. These papers point out that some reports, especially low-quality reports, still require the involvement of human testers to manually inspect and reproduce bugs. Therefore, identifying the report quality is an important prerequisite for automated bug report reproduction.
Chaparro \etal \cite{chaparro2019assessing} propose Euler, which can automatically identify and assesses the quality of the steps to reproduce in a bug report. Euler generates S2Rs from natural language steps and matches the S2Rs to program states and GUI-level app interactions. Different from bug report assessment techniques, which focus on the reproduction steps in test reports, \toolname focuses on the consistency of app screenshots and textual descriptions, which are more straightforward in reflecting the crowdsourced test report quality when matched with app screenshots.\toolname can promote the automated report reproduction process.
}

\subsection{GUI Understanding}

Image understanding for software testing has been an emerging trend. Mobile apps are GUI intensive \cite{yu2021layout}, and many problems would be reflected in GUIs. Therefore, it is significant and effective to start from the understanding of app GUI to assist in software testing. Therefore, GUI understanding technologies have been adopted in many different software testing tasks, and have achieved preliminary success in some areas.
Nguyen \etal \cite{Nguyen2016Reverse} introduced REMAUI, which uses CV algorithms to recognize widgets, texts, and images in GUI screenshots. 
On the basis of REMAUI, Moran \etal \cite{moran2018machine} proposed REDRAW. The approach utilizes machine learning algorithms and can identify the widgets to generate GUI code automatically. 
Similarly, Chen \etal \cite{chen2018ui} also proposed an approach to generate GUI skeletons from UI images. Such an approach combines traditional CV algorithms and ML models.
Xiao \etal \cite{xiao2019iconintent} proposed IconIntent to infer the intents of GUI widgets with CT technologies, and then detect the violations of collecting sensitive data.
Chen \etal \cite{chen2020unblind} introduced an encoder-decoder DL model to generate captions for specific widget images.
Liu \etal \cite{liu2020owl} proposed OwlEye, a tool that adopts deep learning models to detect and locate display issues on GUI images by modeling the visual information.
The above work requires the detection and location of GUI widgets. Chen \etal \cite{chen2020object} conducted a survey on the application of object detection algorithms in GUI widget detection, and proposed a novel algorithm, combining traditional CV and DL models, to improve the widget detection accuracy.
Yu \etal \cite{yu2021layout} \cite{yu2019lirat} proposed an approach named LIRAT to record and replay mobile app test scripts among different platforms with a thorough understanding of the app screenshot and widget layout.
Xu \etal \cite{xu2021guider} proposed Guider, which is used to repair GUI test scripts for Android apps automatically, with the utilization of visual and structural information of widgets to understand what widgets have been updated.
Cooper \etal \cite{cooper2021takes} proposed Tango, which leverages both visual and textual information to detect duplicate video-based reports.
Liu \etal propose Nighthawk \cite{liu2022nighthawk} to have a deep and thorough study on the automated localization of GUI display issues via visual element understanding.
\vv{Feng \etal propose GIFdroid \cite{feng2022gifdroid} to help reproduce the bug reports in the GIF form with image understanding and dynamic app analysis.}
Liu \etal propose NaviDroid \cite{liu2022guided} to assist manual testing via hints extracted from the dynamic analysis of apps.
\vvvvv{Regarding the emerging large language models (LLM), though we cannot evaluate their capability in mobile app GUI understanding, which is quite different from the common life scenario image understanding. LLMs are expected to have a better understanding on the widgets and their layout relationships than normal DL models. We are optimistic about the future application of the LLMs in crowdsourced test report consistency detection.}

Such approaches view apps from the visual perspective. We adapt and combine these visual analysis algorithms to understand the app screenshots in crowdsourced test reports. Then, the analyzed results can be perceived as a reference to detect the crowdsourced test report consistency.

\section{Conclusion}
\label{sec:con}

Crowdsourced testing has been facing severe quality control problems for a long time. Crowdsourced test reports are submitted by crowdworkers of different expertise and are of a wide range of quality. Reviewing such reports is a great waste of time. Therefore, this paper introduces \toolname to detect mobile app crowdsourced test report consistency via deep image-and-text fusion understanding. \toolname is a two-stage approach. First, we classify the reports into different categories according to bug features. Second, based on the classifying results, we design different strategies for different types of bugs according to the deep image-and-text fusion understanding. With the analysis of GUI features, \toolname detects consistency in crowdsourced test reports. Empirical evaluation results show that \toolname has an excellent performance on detecting consistency, and outperforms the state-of-the-art one-stage approach by 116.98\% in accuracy, and outperforms the state-of-the-art bug classification approach by 99.8\%. \vv{Besides, we conduct a user study to evaluate the practical value of \toolname. It shows that \toolname can effectively help developers improve the test report reviewing efficiency.}

\bibliographystyle{IEEEtran}
\bibliography{main}

\end{document}